\documentclass[fleqn,usenatbib]{mnras}

\usepackage{newtxtext,newtxmath}

\usepackage[T1]{fontenc}

\DeclareRobustCommand{\VAN}[3]{#2}
\let\VANthebibliography\thebibliography
\def\thebibliography{\DeclareRobustCommand{\VAN}[3]{##3}\VANthebibliography}

\usepackage{graphicx}	
\usepackage{amsmath}	
\usepackage{xcolor}

\title[ICL in a Fossil cluster]{Intracluster Light Properties in a Fossil Cluster at z=0.47}

\author[Yoo et al.]{
Jaewon Yoo,$^{1,2}$\thanks{E-mail: jwyoo@kasi.re.kr}
Jongwan Ko,$^{1,2}$
Jae-Woo Kim$^{1}$
Hyowon Kim$^{1,2}$
\\
$^{1}$Korea Astronomy and Space Science Institute (KASI), Daedeokdae-ro, Daejeon 34055, Korea\\
$^{2}$University of Science and Technology (UST), Gajeong-ro, Daejeon 34113, Korea
}

\date{Accepted XXX. Received YYY; in original form ZZZ}

\pubyear{2021}

\begin{document}
\label{firstpage}
\pagerange{\pageref{firstpage}--\pageref{lastpage}}
\maketitle

\begin{abstract}
Galaxy clusters contain a diffuse stellar component outside the cluster's galaxies\textcolor{black}{, which} is observed as faint intracluster light (ICL). 
Using Gemini/GMOS-N deep imaging and multi-object spectroscopy of a massive fossil cluster at a redshift of $z=0.47$, RX J105453.3+552102 (J1054), we improve \textcolor{black}{the} observational constraints on the formation mechanism of the ICL.
We extract the ICL surface brightness and colour profiles out to 155 kpc from the brightest cluster galaxy (BCG)
with \textcolor{black}{a} detection limit of 28.7 mag/arcsec$^2$ (1$\sigma$, $4.8\arcsec \times 4.8\arcsec$; \textit{i}--band). 
The colour of the diffuse light is similar to that of the BCG and central bright galaxies out to $\sim$ 70 kpc\textcolor{black}{, becoming slightly} bluer toward the outside. 
We find that the ICL distribution \textcolor{black}{show}\textcolor{black}{s} better \textcolor{black}{agreement} with the spatial distribution of member galaxies than with the BCG-dominated cluster luminosity distribution. 
We report the ICL fraction of J1054 as $15.07 \pm 4.57 \%$ in the range of $60 \sim 155$ kpc from the BCG, which \textcolor{black}{appears to be} higher than the ICL fraction-redshift trend in previous studies. 
Our findings suggest that intracluster stars seems not to be explained by one dominant production mechanism. However, a significant fraction of the ICL of J1054 \textcolor{black}{may} have been generated from the outskirts of infalling/satellite galaxies \textcolor{black}{more recently} rather than \textcolor{black}{by} the BCG at the early stage of the cluster.
\end{abstract}




\section{Introduction}

Recent deep observations of nearby galaxy clusters show distinct diffuse \textit{intracluster light (ICL)}, \textcolor{black}{referring to} the light from stars that are not bound to any individual cluster galaxy \citep{1951PASP...63...61Z, 1998Natur.396..549G, 2002ApJ...575..779F, 2004ApJ...617..879L, 2005ApJ...618..195G, 2005MNRAS.358..949Z, 2005ApJ...631L..41M, 2017ApJ...834...16M, 2018MNRAS.474.3009D, 2019A&A...622A.183J, 2020ApJS..247...43K, 2021MNRAS.502.2419F, 2021ApJ...910...45M}.
\textcolor{black}{The} ICL is expected to follow the global potential well of the galaxy cluster \textcolor{black}{and can therefore} be used as a luminous tracer for dark matter \citep{2018MNRAS.474..917M, 2019MNRAS.482.2838M, 2020MNRAS.494.1859A}. Furthermore, accounting for \textcolor{black}{the} ICL could \textcolor{black}{bolster our} understanding \textcolor{black}{of} the baryon component of the universe \citep{2016ApJ...826..146B} and reduc\textcolor{black}{e} the discrepancy between observation\textcolor{black}{s} and cosmological simulation\textcolor{black}{s}. In other words, observations of the distribution and abundance of ICL would impose a strong constraint on theoretical models of the evolution of galaxy cluster\textcolor{black}{s}  \citep{2004ApJ...617..879L, 2010HiA....15...97A}.   

The abundance \textcolor{black}{of} ICL increases through numerous galaxy interactions in galaxy cluster\textcolor{black}{s}, as suggested by simulation studies \citep{2007MNRAS.377....2M, 2007ApJ...666...20P, 2007ApJ...668..826C, 2010MNRAS.406..936P, 2011ApJ...732...48R, 2014MNRAS.437.3787C, 2015MNRAS.451.2703C}. Observations also reported several scenarios for the production of intracluster stars, including brightest cluster galaxies (BCG) major mergers and violent relaxation afterwards \citep{2007ApJ...665L...9R, 2018ApJ...862...95K}, tidal stripping from the outskirt\textcolor{black}{s} of L$^{\ast}$ member galaxies \citep{2017ApJ...851...75I, 2018MNRAS.474.3009D, 2018MNRAS.474..917M}, disruption\textcolor{black}{s} of dwarf galaxies as they fall towards galaxy cluster centre\textcolor{black}{s} \citep{2011MNRAS.414..602T} and in-situ star formation \citep{2002ApJ...580L.121G}. However, the main ICL formation mechanism is still under debate due to \textcolor{black}{a} lack of ICL measurements over a wide range of redshifts and masses with an equal observational method of measuring the ICL properties.

Within th\textcolor{black}{is} paradigm, \textcolor{black}{whereby} a significant fraction of intracluster stars ha\textcolor{black}{s} been stripped from cluster galaxies through multiple mechanisms, we \textcolor{black}{can} expect that the amount of ICL (\textit{\textcolor{black}{the} ICL fraction}\textcolor{black}{; diffuse light to cluster total light fraction}) can act as a measurement tool for estimating the dynamic state\textcolor{black}{s} of galaxy clusters \citep{2016IAUS..317...27M, 2019arXiv191201616M}. For example, more evolved clusters should have a higher ICL fraction compared to dynamic young clusters. \textcolor{black}{Values in p}revious studies \textcolor{black}{of} ICL fraction\textcolor{black}{s} vary immensely, from 2.6\% \citep{2010MNRAS.403L..79M} to 50\% \citep{2004ApJ...617..879L}, largely because the observed ICL fraction varies according to \textcolor{black}{the} cluster's dynamical state. In this context, important constraints thus come from ICL measurements in the most evolved (dynamically relaxed) galaxy clusters.

According to the current Lambda cold dark matter ($\Lambda$CDM) model, viriali\textcolor{black}{s}ed halos grow by hierarchical merging. \textit{Fossil clusters} \citep{1994Natur.369..462P, 2003MNRAS.343..627J, 2006AJ....132..514C} are galaxy clusters in which the accretion rate \textcolor{black}{i}s fast and efficient, resulting \textcolor{black}{in} luminosity-dominant BCGs \citep{2005ApJ...630L.109D}. Fossil clusters are defined as having an absolute magnitude gap in \textcolor{black}{the} \textit{r}--band between the BCG and the second brightest galaxy ($\Delta M_{12}$) \textcolor{black}{that exceeds} 2, containing extended X-ray sources with $L_X > 10^{42}erg/s$ \citep{2003MNRAS.343..627J}. In fossil clusters, all \textcolor{black}{significant merging and} violent relaxation events have already taken place, thus we regard them as dynamically evolved and relaxed systems. Because \textcolor{black}{they may have undergone significant dynamic processes which enrich both the ICL and the BCG \citep{2014MNRAS.437.3787C, 2019ApJ...871...24C}}, fossil clusters are expected to have the most abundant ICL fraction among galaxy clusters \textcolor{black}{with} similar redshifts and masses. Moreover, the dominant BCGs of fossil clusters allow us to assume a common assembly history, which is \textit{``hierarchical merging toward one central mass."} 

Measuring the ICL fraction in fossil clusters at various redshift\textcolor{black}{s} will thus help us improve our understanding about ICL formation and evolution. However, high-z samples are very limited, \textcolor{black}{as} fossil clusters are themselves rare \citep{2003MNRAS.343..627J, 2011PASP..123....1T} and there \textcolor{black}{is too little} spectroscopic membership data to fulfil the definition of \textcolor{black}{a} fossil cluster. The detection of ICL in high-z fossil clusters is even more challenging due to its faint\textcolor{black}{nes}s. Although several intermediate-z fossil clusters have been discovered \citep{2005ApJ...624..124U, 2011MNRAS.417.2927P}, ICL has not been studied in detail in \textcolor{black}{relation to} fossil clusters. The only fossil cluster for which there has been an ICL study is Abell 2261 at a redshift of $\sim$ 0.22 with \textcolor{black}{an} ICL fraction $\sim$ 16.64\% \citep{2015MNRAS.449.2353B}.

There is a consensus that most intracluster stars are produced after $z \sim 1$ \citep[e.g.,][]{2010MNRAS.406..936P, 2011ApJ...732...48R, 2015MNRAS.451.2703C}. Interestingly, some studies \citep{2014MNRAS.437.3787C, 2015MNRAS.449.2353B, 2017ApJ...851..139L} claim that the majority of \textcolor{black}{the} stellar mass of BCGs ha\textcolor{black}{s} assembled by $z \sim 0.4$ and \textcolor{black}{that} the accreted stellar mass after this epoch mainly contribute\textcolor{black}{s} to the growth of the present-day ICL. If BCG growth is the main ICL formation mechanism, then the ICL fraction at $z \sim 0.4-0.5$ is likely to be as high as \textcolor{black}{that of} local galaxy clusters and the colour of the ICL is likely to be similar to that of the BCG (optically red). If tidal stripping/disruption of infalling/satellite galaxies is the main mechanism, the ICL fraction is likely to show a strong evolution from $z \sim 0.4-0.5$, and its colour would be much bluer than BCG and galaxies near the cluster core because stripped stars mainly originate from \textcolor{black}{the} outskirts of galaxies with low metallicity or blue dwarf galaxies with young stellar populations. Thus, fossil clusters around $z \sim 0.4-0.5$ can be an optimal target to examine the BCG-related ICL formation scenario.

Here, we report measurements of \textcolor{black}{the} ICL distribution, colour, and fraction \textcolor{black}{relative} to the total cluster light using deep optical imaging and multi-object spectroscopy data of a fossil cluster at $z=0.47$.
To date, this is the most distant fossil cluster for which the distribution and abundance of ICL \textcolor{black}{have been} measured.

This paper has \textcolor{black}{the} following structure. We describe the details of our observations and the data reduction procedure in Section 2. We analy\textcolor{black}{s}e the robustness of \textcolor{black}{the} ICL measurement \textcolor{black}{scheme} in Section 3. In Section 4, we report the results o\textcolor{black}{f} the ICL measurements and galaxy cluster properties. The possible ICL formation scenario based on the result\textcolor{black}{s} is discussed in Section 5. Finally, we summari\textcolor{black}{s}e and conclude \textcolor{black}{the paper} in Section 6. Throughout the paper, we adopt a standard $\Lambda$CDM model with $H_0$=70 km s$^{-1}$ Mpc$^{-1}$, $\Omega_m$=0.3 and $\Omega_\Lambda$=0.7, \textcolor{black}{providing an} angular scale of 5.89 kpc arcsec$^{-1}$ at \textcolor{black}{a} redshift of \textcolor{black}{the target cluster}. Magnitudes are \textcolor{black}{expressed} in the AB magnitude system.

\section{Data} \label{sec:obs}

We conducted deep imaging observations and multi-object spectroscopy (MOS) of the fossil cluster at $z=0.47$ using \textcolor{black}{the} GMOS-N instrument \citep{2004PASP..116..425H} o\textcolor{black}{f} the 8.1 meter Gemini North telescope. The field of view of GMOS is $5.5' \times 5.5'$, which corresponds to $\sim$ 2.0 Mpc $\times$ 2.0 Mpc at the redshift of the target cluster. The data were obtained \textcolor{black}{i}n April and May \textcolor{black}{of} 2018 (GN-2018A-Q-201/PI: Jaewon Yoo).
Table \ref{tab:mathmode2} describes our observations, including \textcolor{black}{the} date, seeing conditions, filters, the number of exposures, and \textcolor{black}{the} exposure time per visit. 
Using the \texttt{Gemini IRAF\footnote{IRAF is distributed by the National Optical Astronomy Observatory, which is operated by the Association of Universities for Research in Astronomy(AURA) under cooperative agreement with the National Science Foundation.} package}, we performed basic data reduction following the \texttt{Gemini data reduction cookbook} \citep{2016gemini}.

\subsection{Target \textcolor{black}{c}luster \textcolor{black}{s}election}

We selected our target galaxy cluster from the fossil system candidates \citep{2007AJ....134.1551S}, which are based on \textcolor{black}{the} SDSS DR5 \citep{2007ApJS..172..634A} and ROSAT \citep{1999A&A...349..389V} catalog\textcolor{black}{ue}s.
Among the 34 fossil system candidates in Santos et al., we \textcolor{black}{initially} selected massive galaxy clusters using an X-ray luminosity cut, i.e. $L_X \geq 10^{44}erg/s$. In the SDSS DR14 \citep{2018ApJS..235...42A} catalog\textcolor{black}{ue}, we retrieved potential member galaxies (within a projected radius of 500 kpc from the BCG and with a photometric redshift cut $z_{BCG} \pm  0.05$) of the galaxy clusters. The photometric redshift is based on the \texttt{kd-tree nearest neighbor fit} method \citep{2007AN....328..852C}.
In the redshift range of $0.4 < z < 0.5$, we selected RX J105453.3+552102 (hereafter J1054), which is one of the fossil group origins (FOGO) project sample\textcolor{black}{s that} has been confirmed as a fossil cluster \textcolor{black}{with $\Delta M_{12} = 2.12 \pm 0.33$} \citep{2011A&A...527A.143A, 2014A&A...565A.116Z}. Table \ref{tab:target_table} lists \textcolor{black}{the} key properties of J1054. 

The presence of the galactic cirrus can lead to misinterpretation\textcolor{black}{s} in ICL measurements. \textcolor{black}{For this reason,} we checked the Planck 2015 dust map \citep{2016A&A...594A..10P}. The target cluster field has no extended emission. Additionally, we found E(\textit{B}--\textit{V})=$0.0088 \pm 0.0003 $ for the target field from another dust map (based on Gaia, Pan-STARRS 1, and 2MASS) provided by \cite{2019ApJ...887...93G}.

\begin{table*}
\centering
\caption{Target description}
\label{tab:target_table}
\begin{tabular}{cccccccc}
\hline
SDSS name & ROSAT name & R.A & Dec & z & $L_X$ &$r_{200}^a$ & $M_{200}^a$ \\
 &  & (h:m:s) & (d:m:s) &  & [$10^{44}erg/s$] &[Mpc] & [$10^{14} M_{\odot}$] \\
\hline
J105452.03+552112.5 & RX J105453.3+552102 & 10:54:52.03 & +55:21:12.5 & 0.47 & 2.39 & 1.68 & 5.40  \\
\hline
\multicolumn{8}{l}{$^a$ The $r_{200}$ and the $M_{200}$ are calculated in Section 4.2.2.}\\
\end{tabular}
\end{table*}

\subsection{Observations and \textcolor{black}{r}eductions}

\begin{table*}
\centering
\caption{GMOS-N observation}
\label{tab:mathmode2}
\begin{tabular}{cccccccc}
\hline
Observation Date & \textcolor{black}{Days from}& Seeing$^b$ & Moffat $\beta$$^b$ & Filter & Number of exposure & Exposure time per visit & Used in science \\
(YYYY-mm-dd) & \textcolor{black}{New Moon}& [arcsec] & & &  & [s] & \\
\hline
2018-04-11   & \textcolor{black}{25}& 0.97\arcsec & 2.65 &\textit{i} & 24 & 178 & Yes\\
2018-05-19   & \textcolor{black}{4}& - & - & \textit{r} & 6 & 170 & No\\
2018-05-21   & \textcolor{black}{6}& - & - & \textit{i} & 4 & 178 & No\\
2018-05-21   & \textcolor{black}{6}& 0.76\arcsec & 3.06 & \textit{r} & 19 & 170 & Yes\\
2018-05-21   & \textcolor{black}{6}& - & - & \textit{OG515} (MOS) & 3 & 600 & Yes\\
\hline
\multicolumn{7}{l}{$^b$ Point spread function (PSF) of each band image is modeled using the \texttt{PSFex} \citep{2011ASPC..442..435B} routine.}\\
\end{tabular}
\end{table*}


Measurements of the ICL demand careful observations and data reduction \citep{2005ApJ...631L..41M, 2015MNRAS.446..120D, 2015A&A...581A..10C, 2016ApJ...823..123T, 2019A&A...621A.133B, 2020A&A...644A..42R, 2021ApJ...910...45M}. 
 In general, the key \textcolor{black}{during} ICL measurements is to minimi\textcolor{black}{s}e \textcolor{black}{the level of} sky background error, mostly dominated by residual flat-fielding inaccurac\textcolor{black}{ies} on a large scale and sky background subtraction. 
We \textcolor{black}{describe below} \textcolor{black}{the} observational strategy and data processing \textcolor{black}{method} in detail to achieve our goal in this study.

\begin{figure*}
\centering
\includegraphics[width=2.0\columnwidth,trim={0.5cm 1.5cm 1cm 1.5cm},clip]{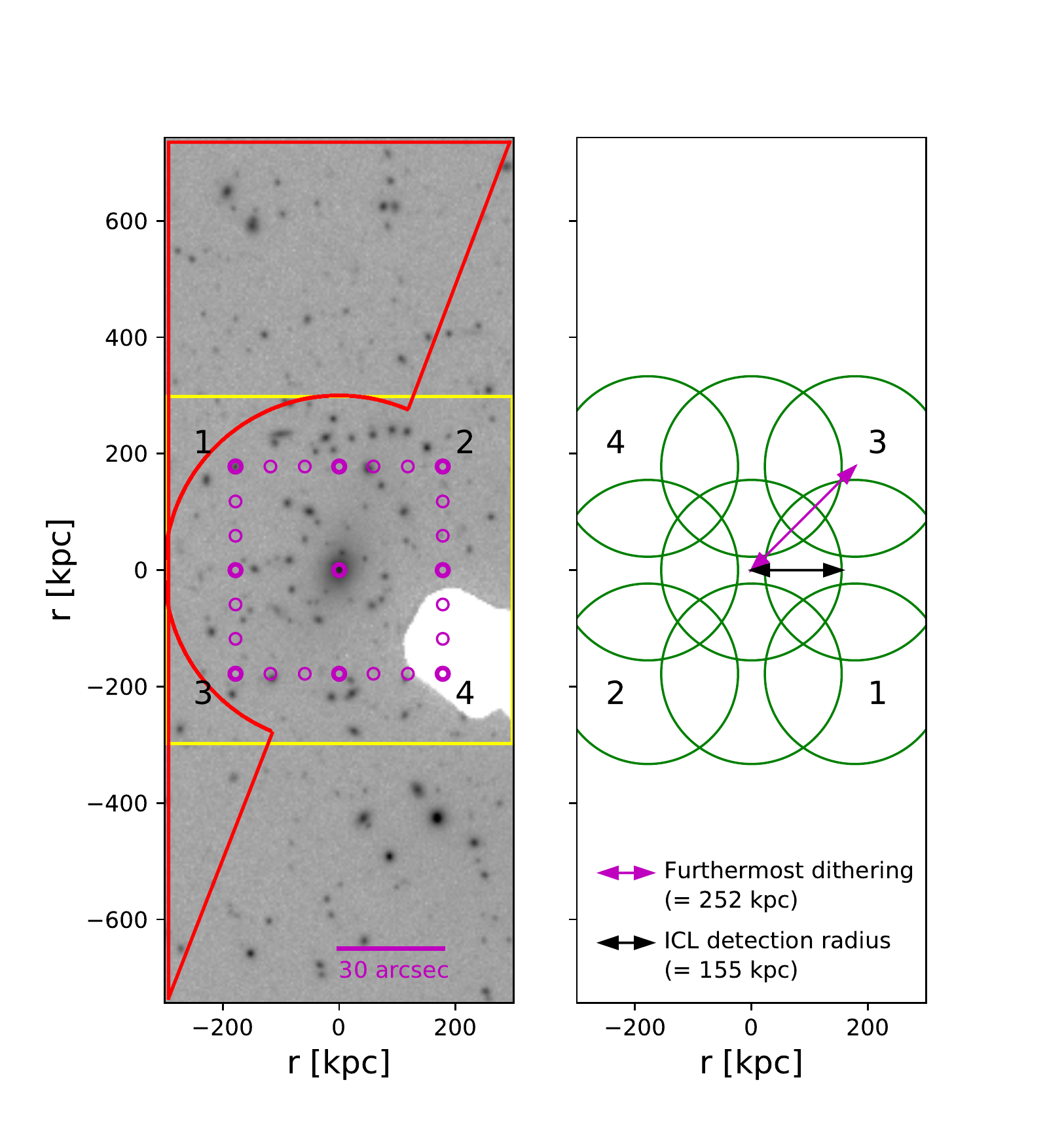}
\caption{(Left) The \textit{i}--band deep image of \textcolor{black}{the} J1054 cluster. \textcolor{black}{Here,} 24 exposure frames are combined and \textcolor{black}{the corresponding} full exposure area is shown. The scales are inverse and log-scaled. The dithering patterns are \textcolor{black}{denoted by the} magenta circles. The furthermost dither is $42.5''$ ($\sim$ 252 kpc) away from the centre. The white region \textcolor{black}{toward the} south is where the guide arm \textcolor{black}{is} locate\textcolor{black}{d}. \textcolor{black}{The background level and \textcolor{black}{corresponding} error are measured in the red defined region.} (Right) \textcolor{black}{Animation of} the dithering strategy. From the \textcolor{black}{nine} dithering points (\textcolor{black}{indicated by the} thick magenta circles in the panel \textcolor{black}{on the left}), \textcolor{black}{predi}cted ICL regions in the sky flat are \textcolor{black}{indicated by the} green circles. The numbers 1 - 4 indicate the ICL region corresponding to the dithering position with the same numbering in the left panel. See the text for more detail\textcolor{black}{s about the} observation strategy and \textcolor{black}{the} masking \textcolor{black}{done} for \textcolor{black}{the} analysis.
\label{fig:f1}}
\end{figure*}

\textit{\bf{Observation strategy}}: In the field of the cluster, there is a bright foreground star (\texttt{cModelMag} $m_i$=14.58 mag) which imposes a strong limit on the exposure time. To reach \textcolor{black}{a} low surface brightness regime, we decided to use \textcolor{black}{this} as a guide star \textcolor{black}{to} be covered by \textcolor{black}{a} guiding arm, thus securing an appropriate exposure time and \textcolor{black}{reducing} overhead. 
We took dithering steps large (furthermost dithering from the centre is 42.5\arcsec; see Figure \ref{fig:f1}) enough to \textcolor{black}{prevent} the central region of the cluster \textcolor{black}{from} being overlapped between exposures and to obtain a flat background around the cluster ICL region. This also allows us to make sky flats, removing the area vignetted by the guiding arm.


\textcolor{black}{To} measure \textcolor{black}{the} colour of \textcolor{black}{the} ICL and cluster member galaxies, we decided to observe in two \textcolor{black}{bands, the} passband \textit{i}-- and \textit{r}--band\textcolor{black}{s, as they allow}, detect\textcolor{black}{ion of} the red sequence using the 4000\AA\ break and ha\textcolor{black}{ve} good instrument sensitivity. 
After blocking the brightest star, to \textcolor{black}{prevent} saturation \textcolor{black}{by other} bright stars \textcolor{black}{from} affecting the ICL analysis, the single exposure time was designed to be 178s and 170s for \textcolor{black}{the} \textit{i}--band and \textit{r}--band, respectively.
The observed exposure time is 83 minutes for the \textit{i}--band and  71 minutes for the \textit{r}--band (see Table \ref{tab:mathmode2}).
The original field of view of GMOS is $5.5' \times 5.5'$\textcolor{black}{;} nonetheless, we decided to use only part of the image \textcolor{black}{given that} our analysis is extremely sensitive to background noise. After dithering and avoiding chip gaps, the final image area covered by every exposure has a size of $1.7' \times 4.0'$. This \textcolor{black}{area of} full exposure corresponds to 600 kpc $\times$ 1415 kpc at the target redshift \textcolor{black}{with} a radius of 300 kpc from \textcolor{black}{the} BCG centre.

\begin{figure*}
\centering
\includegraphics[width=2.0\columnwidth,trim={0 12cm 0 10cm},clip]{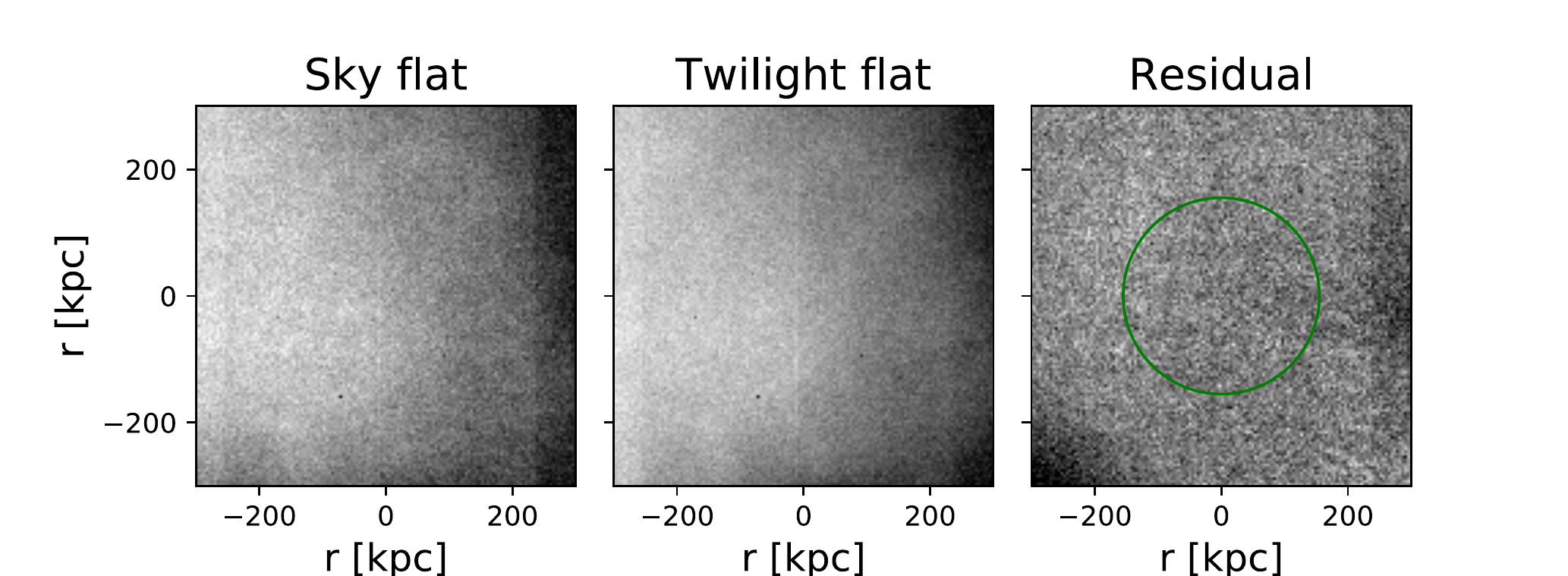}
\caption{(Left) The \textit{i}--band sky flat generated by combining 24 science images. The central 600 kpc $\times$ 600 kpc region (\textcolor{black}{indicated by the} yellow box in Figure \ref{fig:f1}) of the cluster, where we measure \textcolor{black}{the} ICL, is plotted. (Middle) For the same region, the \textit{i}--band twilight flat generated by combining 34 twilight images. (Right) The residual (sky flat -- twilight flat) image. Its standard deviation is 0.003. The green circle indicates the central region with \textcolor{black}{a} radius of 155 kpc, \textcolor{black}{within which} the ICL analysis is conducted. The scales are z-scaled and the scale bar of each image is plotted on the bottom side. The residual image shows that the sky flat does not have art\textcolor{black}{e}facts from bright objects or \textcolor{black}{from} the guide arm \textcolor{black}{(see the panel on the left in Figure \ref{fig:f1} for their location)}.  
\label{fig:flat}}
\end{figure*}

\textit{\bf Flat-fielding}: \textcolor{black}{F}lat-fielding is a crucial step in the data reduction \textcolor{black}{process} for low surface\textcolor{black}{-}brightness science. To minimi\textcolor{black}{s}e inaccurac\textcolor{black}{ies during} flat-fielding, we created sky flats using our science images taken on the same night (i.e., 24 and 19 images taken \textcolor{black}{during} one night for \textcolor{black}{the} \textit{i}-- and \textit{r}--band\textcolor{black}{s}, respectively). For each science image, we masked out objects (\textcolor{black}{more than} 3$\sigma$ from the median) over two iterations and normalised \textcolor{black}{each} image using the mode value. \textcolor{black}{The} masked and scaled input images were \textcolor{black}{then} median\textcolor{black}{-}combined to form a single sky flat.

The large offsets between exposures (for each exposure, 19 exposures are at least 178 kpc away) allow the central region near the BCG, where we measure \textcolor{black}{the} ICL and cluster galaxies to \textcolor{black}{prevent them as much as possible from} occupying the same physical region of the detector in all science exposures. 
In Figure \ref{fig:flat}, we show the difference between the sky and twilight flats in \textcolor{black}{the} \textit{i}--band. The standard deviation of the residual image is 0.003.
Within the central region where we measure \textcolor{black}{the} ICL (the green circle in Figure \ref{fig:flat}), our sky flat does not show any trend of higher pixel values compared to the twilight flat, which is free from central diffuse light and the guiding arm. 
Thus, our sky flat \textcolor{black}{is} not \textcolor{black}{affected by} any gradient caused by diffuse light in the cluster central region. 
\textcolor{black}{Moreover,} artifacts from the guide arms are naturally removed when the images \textcolor{black}{are stacked}. The absence of any suspicious residual light \textcolor{black}{possibly affected by} the guide arm furthermore ensured us that the effect of the guide arm \textcolor{black}{on} our sky flat is negligible.

\textit{\bf Astrometry refinement}: 
For the refinement of the astrometry \textcolor{black}{of} all \textcolor{black}{of} our science images, we used the \texttt{mscfinder.msctpeak} task in \texttt{IRAF}, as suggested \textcolor{black}{in} the \texttt{Gemini data reduction cookbook}\footnote{http://ast.noao.edu/sites/default/files/GMOS\_Cookbook/Processing/IrafPr ocImg.html\#wcs-refinement}. This task allowed us to characteri\textcolor{black}{s}e the distortions through the TNX projection, which is an experimental tangent plane projection method. We take as a reference the astrometry of the stars in the SDSS DR14 catalog\textcolor{black}{ue}. To do this, we selected SDSS stars across the image, satisfying our point source criteria, \texttt{psfMag\_i}-\texttt{cModelMag\_i} $<0.2$ and non-saturated 20 $<$ \texttt{cModelMag\_i} $<$ 22. The \texttt{cModelMag} \textcolor{black}{value} is derived from a linear combination of the exponential fit and the de Vaucouleurs fit, and a perfect point source would have \textcolor{black}{an} exponentially decaying radial profile. Therefore, the \texttt{cModelMag} and point spread fun\textcolor{black}{c}tion fitted \texttt{psfMag} \textcolor{black}{values} have excellent agreement for stars \citep{2004AJ....128..502A}.
Through th\textcolor{black}{e}se criteria, 50 stars were selected and \textcolor{black}{approximately} 40 stars were matched in each image. 
We found \textcolor{black}{that} the astrometry precision is about $ 0.02\arcsec$.

\textit{\bf Sky subtraction}: Another critical step is consistent sky subtraction for all science exposures before image stacking. 
We determine \textcolor{black}{the} global sky level for each exposure by estimating the median of the statistical distribution of pixels in the image with sigma-clipping over \textcolor{black}{five} iterations using \textcolor{black}{the} \texttt{imstatistics} task in \texttt{IRAF} and subtract this value from \textcolor{black}{the} individual images. 
If the ICL is no\textcolor{black}{n-}negligible on a large scale (in the central region near the BCG), the sky level is likely to be overestimated and the ICL will \textcolor{black}{therefore} be underestimated. However, we do not have to worry about this removal of \textcolor{black}{the} ICL because we later separately measure the sky background level again from the final stacked image. Above this background level, the ICL is quantified by measuring the excess surface brightness. 


\textit{\bf Image coaddition}: Final stacking of the individual science images was performed using the \texttt{imcoadd} task in \texttt{IRAF}. After removing artefacts (bad pixels, cosmic rays, satellite tracks) individual images were median\textcolor{black}{-}combined. The image resampling method used here is \texttt{nearest}. Figure \ref{fig:f1} shows the full exposure area in the \textit{i}--band.

\textit{\bf PSF matching}: The reduced \textit{i}--band images were taken in slightly worse seeing conditions than the \textcolor{black}{in} \textit{r}--band (see Table \ref{tab:mathmode2}). Even if we use a common mask for the \textit{i}-- and \textit{r}--band\textcolor{black}{s} (presented in Section 3.2), different PSF profiles could cause the ICL measurements to differ \citep{2016ApJ...823..123T, 2017A&A...601A..86K}. To address this issue, it is necessary to degrade the \textit{r}--band images (good seeing; 0.76$\arcsec$)  down to the seeing \textcolor{black}{level} of the \textit{i}--band images (bad seeing; 0.97$\arcsec$). \textcolor{black}{First,} we used \texttt{PSFex} to model the PSFs in the stacked images for the \textit{i}-- and \textit{r}--band. \texttt{PSFex} gives us a fitted Moffat profile parameter $\beta$ and \textcolor{black}{the} full-width half maximum (FWHM), which determine the shape and scale of the PSF profile. \textcolor{black}{Here,} we adopt the Moffat PSF profile \textcolor{black}{because} it allows a thicker tail compare\textcolor{black}{d} to the Gaussian \textcolor{black}{option} \citep{2001MNRAS.328..977T}. \textcolor{black}{Then, we convolved the \textit{r}--band images using the  \texttt{Moffat2DKernel astropy package} \citep{2013A&A...558A..33A} so as to match the PSF of the \textit{i}--band.} 


\textit{\bf Photometric calibration}: Photometric calibration \textcolor{black}{of} the images is based on SDSS DR15 \citep{2019ApJS..240...23A}, comparing the reference magnitude from the SDSS DR15 photometric catalog\textcolor{black}{ue} to our own.
Due to the lack of bright stars \citep[14.5 $<$ \texttt{psfMag} $<$ 19.5;][]{2008AJ....135..264C}, we calculated a weighted mean of 12 stars within our field of view, \textcolor{black}{assigning} more weight \textcolor{black}{to} brighter stars. 
The estimated zero-point \textcolor{black}{values are} 33.96 mag and 33.93 mag for the \textit{i}-- and \textit{r}--band, respectively. 

\subsection{Spectroscopic data}
The aim of the MOS observation is to determine the redshift\textcolor{black}{s} of potential member galaxies and \textcolor{black}{to} confirm cluster members. Galaxy cluster membership is necessary, not only to confirm the target cluster as a fossil cluster, but also to estimate the total cluster light.
Although the \textcolor{black}{78} spectroscopically confirmed member galaxies were already available in the literature \citep{2011A&A...527A.143A}, we conducted additional MOS observations to identify more member galaxies. We assigned a slit to the BCG to obtain more detailed spectral information, 16 slits for potential member galaxies, \textcolor{black}{five} slits \textcolor{black}{to} obtain a sky spectrum and one slit for a standard star to calibrate \textcolor{black}{the} fluxes. In the results section, we re-selected cluster member galaxies, including new candidate galaxies from the MOS observation.

The potential member galaxies to assign in MOS slits were selected in a \textcolor{black}{manner} similar \textcolor{black}{to that} described in Section 2.1. In the SDSS DR14 catalog\textcolor{black}{ue}, we retrieved galaxies within the GMOS field of view and applied a photometric redshift cut \textcolor{black}{of} $z_{BCG}\pm 0.05$. Then, we excluded galaxies \textcolor{black}{that already} have spectroscopic data. Among the remaining 28 galaxies, considering \textcolor{black}{the} spatial locations of \textcolor{black}{the} slits on the mask and priorit\textcolor{black}{ising} bright and red galaxies, 16 galaxies were assigned via \textcolor{black}{the} \texttt{Gemini MOS Mask Preparation Software (GMMPS)} tool.

We used 1\arcsec\ slit width\textcolor{black}{s} and 5\arcsec\ slit length\textcolor{black}{s}, with the R150 grating and OG515 blocking filter. In each frame \textcolor{black}{two} pixels are binned in both the spectral \textcolor{black}{(0.193 nm/pixel)} and spatial dimensions. Spectra were obtained by combining dithers at \textcolor{black}{the} three central wavelengths of 717, 737, \textcolor{black}{and} 757 nm to fill the gap\textcolor{black}{s} between the detectors. Three exposures \textcolor{black}{of} 600 seconds for each wavelength dithering made the total integration time 30 minutes.

We reduced the MOS data following the official \texttt{Gemini data reduction cookbook} and measured the redshift using \texttt{SpecPro} \citep{2011PASP..123..638M} in \texttt{IDL}.
Among the 16 galaxies in the MOS slits, we \textcolor{black}{could} reliably measure \textcolor{black}{the} redshifts of \textcolor{black}{three} galaxies\textcolor{black}{, whose redshifts estimated to be similar to the BCG redshift. These were therefore included in} the cluster member candidates.



\section{Analysis} \label{sec:analysis}

First, we model \textcolor{black}{the} stars in Section 3.1. In Section\textcolor{black}{s} 3.2 and 3.3, we robustly mask out all luminous objects and determine our background level and \textcolor{black}{the corresponding} error for the ICL measurements. We use only the regions where full exposures cover the stacked image. Additionally, we do not use the regions \textcolor{black}{of} images where the background value could be contaminated by the glow from the guide arm and\textcolor{black}{/or} a bright star with $m_i$=17.20 mag. \\

\subsection{Model\textcolor{black}{l}ing stars} 
Stars have extended stellar wings which could contaminate the colour of the ICL and its fraction \citep{1991ApJ...369...46U} \textcolor{black}{and for this reason} we mask them out completely. We use the \texttt{PSFex} result (presented in \textcolor{black}{the} \textit{PSF matching} part in Section 2.2) to measure the radius of each extended stellar wing. The Moffat profile,  
\begin{equation}
I(r)=I_0\left[1+\left(\frac{r}{\alpha}\right)^2\right]^{-\beta},
\end{equation}

with a scale factor $\alpha=FWHM / 2\sqrt{2^{1/\beta}-1}$,
is more suitable for describing \textcolor{black}{an} extended stellar wing with a thicker tail compare\textcolor{black}{d} to the Gaussian profile.
The resultant Moffat parameter $\beta$ is 2.97 (3.01), \textcolor{black}{and} the FWHM is $ 0.97''$ ($ 0.76''$) for \textcolor{black}{the} \textit{i}--band (\textit{r}--band) image. From the Moffat parameter \textcolor{black}{and} the FWHM of each band image, and \textcolor{black}{after} substituting the peak ADU value of each star for $I_0$ and the background value for $I(r)$, we could calculate the radius of the extended stellar wing of each star analytically.

\subsection{Object detection and \textcolor{black}{m}asking}
Object detection is performed using the software \texttt{SExtractor} \citep{1996A&AS..117..393B}. For rigorous masking, we choose the code parameters through \textcolor{black}{a} visual inspection as \texttt{DETECT\_THRESH} 1 $\sigma$, \texttt{DETECT\_MINAREA} 3 pixel, \texttt{BASK\_SIZE} 64 pixel (corresponding to 60.82 kpc at target redshift), \texttt{DEBLEND\_NTHRESH} 1 (minimum deblending to detect bright source\textcolor{black}{s} strictly and mask out) and \texttt{FILTER\_NAME} gauss\_4.0\_7x7.conv (matching to the seeing on \textcolor{black}{the} observation date). This parameter set is in line with \textcolor{black}{that} adopted by \cite{2018ApJ...862...95K}. Considering the `cold + hot' extraction technique \citep{2004ApJS..152..163R, 2012MNRAS.422..449B, 2013ApJS..206...10G} to detect both extended bright objects and small faint objects, we confirm that our mask (defined by our parameter combination) covers regions of various detection modes \textcolor{black}{in} our analysis region.

Our bright source detection \textcolor{black}{method} is so rigorous that every stellar wing was within the masked region. We also confirm that all stellar wings are outside our ICL analysis region, which is between 60 kpc \textcolor{black}{and} 155 kpc from the centre of \textcolor{black}{the} BCG.
\textcolor{black}{M}asking o\textcolor{black}{f both} bright objects and extended stellar wings \textcolor{black}{is} done in \textcolor{black}{the} \textit{i}--band and applied to both the \textit{i}-- and \textit{r}--band, \textcolor{black}{as} \textit{i}--band images have larger seeing \textcolor{black}{values} and were observed \textcolor{black}{to be} deeper than \textcolor{black}{those of the} \textit{r}--band with longer exposure time\textcolor{black}{s}.

\textcolor{black}{The} bright star and guide arm \textcolor{black}{o}n the south-west side of image in Figure \ref{fig:f1} cause a glow that affects the analysis of diffuse light. After checking the affected area using the \textcolor{black}{corresponding} radial profiles, we decide\textcolor{black}{d} to use only the north-east half of the image for the analysis of the ICL (see the top panel in Figure \ref{fig:f_expMask}). The extended stellar wing of the brightest star \textcolor{black}{measured through the Moffat profile} used for guide star has \textcolor{black}{a} radius of $ 7.61''$. \textcolor{black}{This value} is well within the mask, 
where the distance from the star to the analysis region is $ 35.72''$. 

\begin{figure}
\centering
\includegraphics[width=1.0\columnwidth,trim={0 8cm 0 8cm},clip]{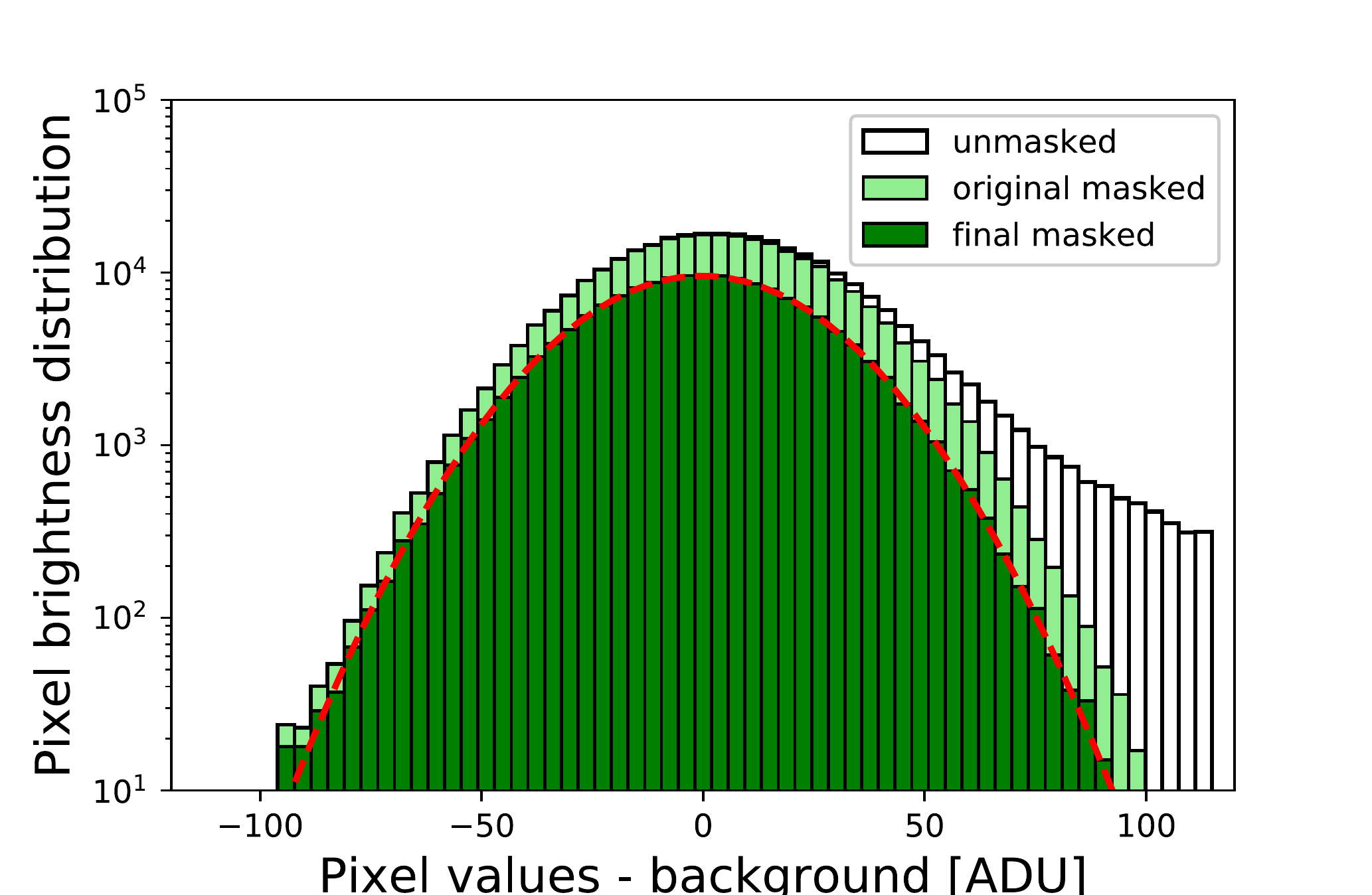}
\caption{\textcolor{black}{D}istributions of background pixel values before masking (white bar), after \textcolor{black}{the} original masking (light green bar) and after expanded masking (green bar) are plotted. After the final masking\textcolor{black}{,} the pixel values show a Gaussian distribution (red dashed line) with improved skewness from 14.47 (before masking) to 0.02 (after original masking) \textcolor{black}{and} to 0.005 (after expanded masking). \label{fig:f3}}
\end{figure}

\begin{figure}
\centering
\includegraphics[angle=90,width=1.0\columnwidth,trim={6cm 2cm 2.5cm 2cm},clip]{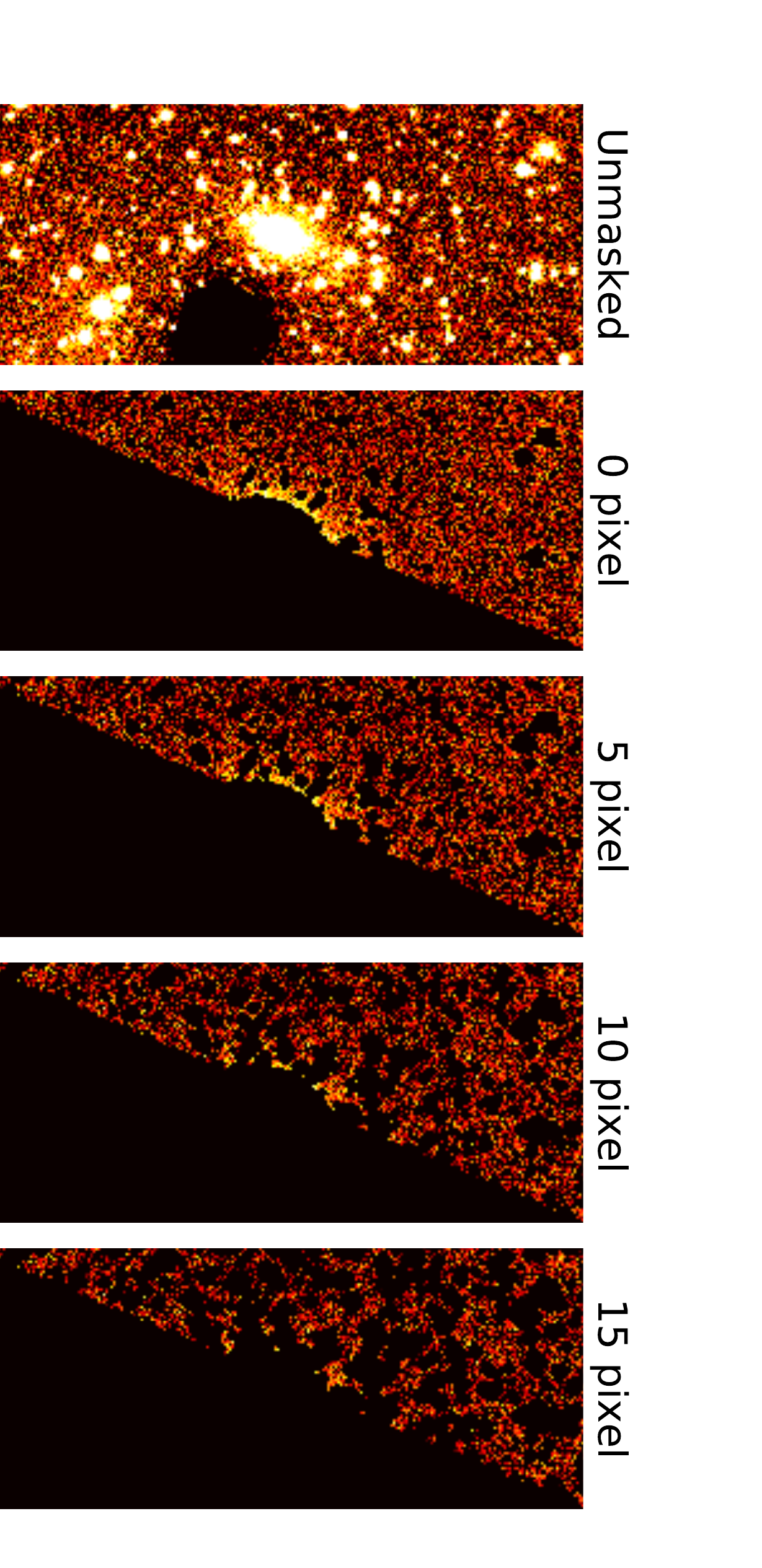}
\includegraphics[angle=90,width=1.0\columnwidth,trim={6.5cm 2cm 8.4cm 2cm},clip]{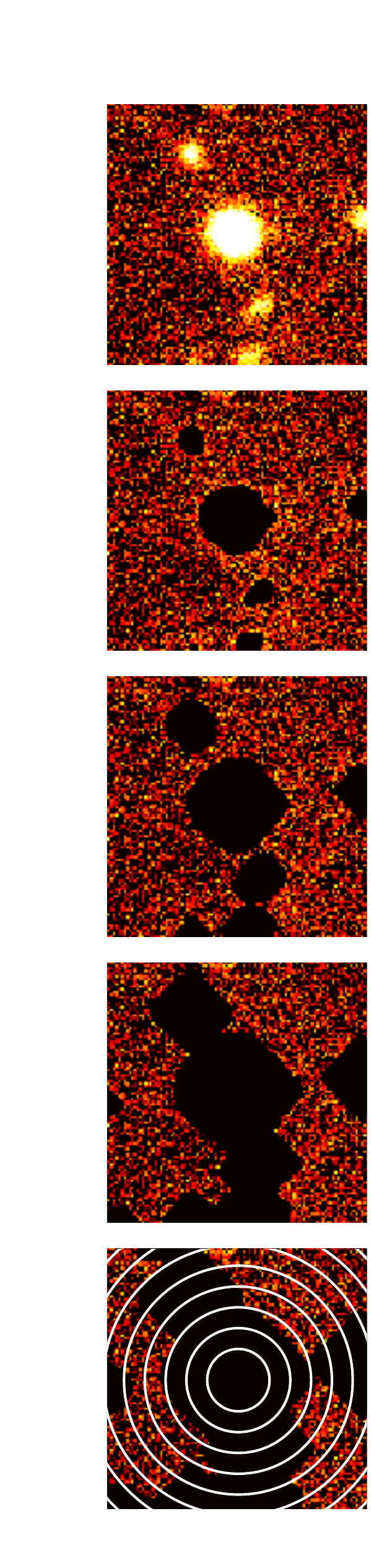}
\includegraphics[width=1.0\columnwidth,trim={0 8.0cm 0 9.5cm},clip]{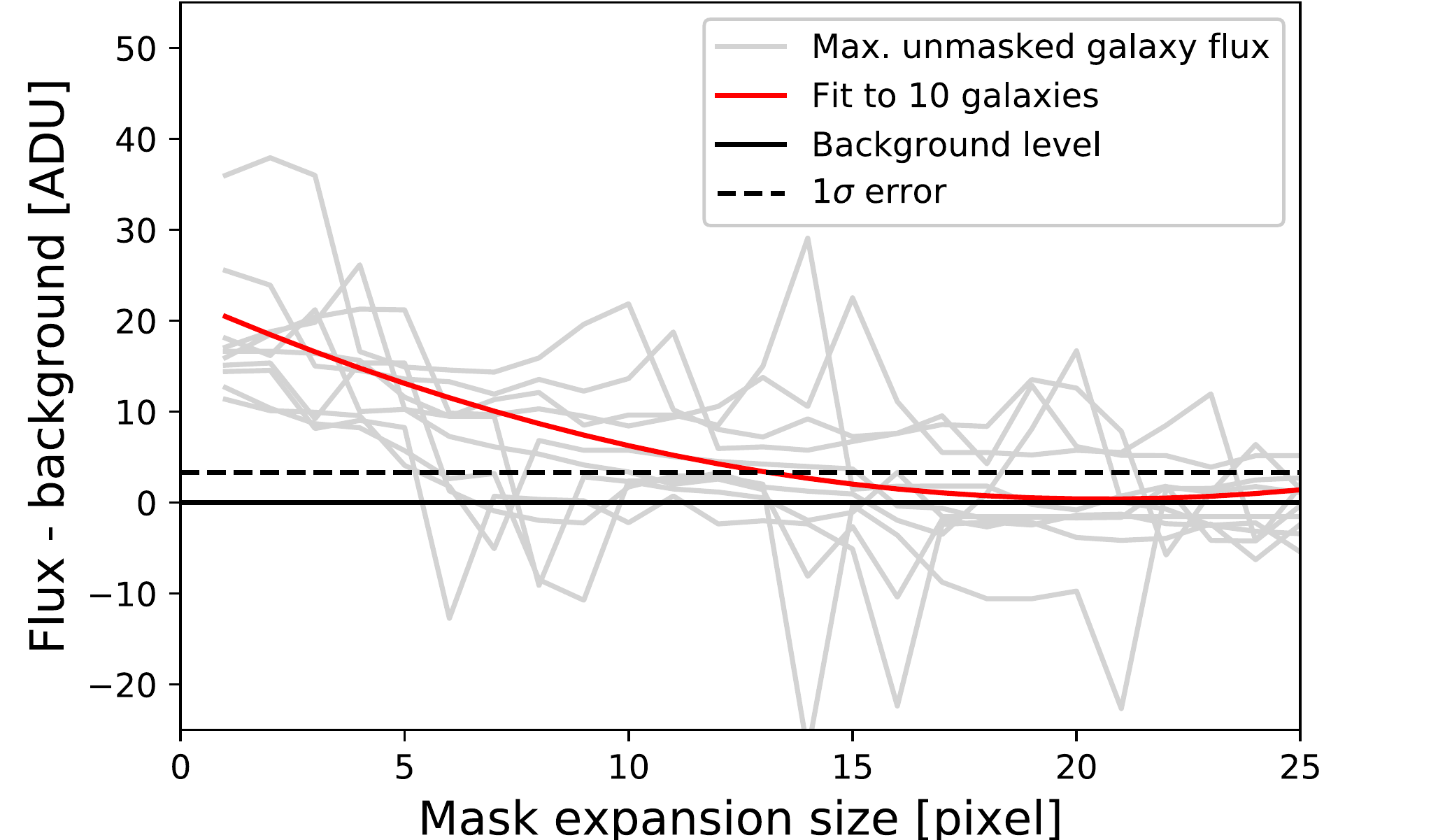}
\caption{(Top) The \textit{i}--band images for each masking step. \textcolor{black}{The corresponding} unmasked, original masked, \textcolor{black}{five} pixel expanded masked, \textcolor{black}{ten} pixel expanded masked and 15\textcolor{black}{-}pixel expanded masked images are shown in a row. (Middle) Stamp images of an example galaxy for each masking step, ordered as in the panel \textcolor{black}{above}. The radial bins \textcolor{black}{used} to measure the \textcolor{black}{profile of the} remaining light are \textcolor{black}{denoted by the} marked as white rings. (\textcolor{black}{Bottom}) Plot of the remaining light as the mask size expands. \textcolor{black}{Ten} galaxies are plotted as gr\textcolor{black}{e}y lines and \textcolor{black}{the corresponding} fit is plotted \textcolor{black}{with the} red line. The fitting line drops below the background error level from the 15\textcolor{black}{-}pixel enlarged mask, \textcolor{black}{corresponding to the final mask size chosen}. See more detail\textcolor{black}{s about this} method in Section 3.2. \label{fig:f_expMask}}
\end{figure}

The reliability of the masking procedure is tested as follows. We draw a radial profile from the centre of the BCG and check the location where it becomes flat, which is 170 kpc from the centre. More safely, we assume that \textcolor{black}{the area beyond} 300 kpc from the centre is free from ICL and regard it as the background region. If the masking procedure is successful, the remaining pixel counts should have a random Gaussian distribution. Figure \ref{fig:f3} shows the pixel count distribution in the background before masking (white) and after masking (light green). After masking, the distribution is well fitted by a Gaussian distribution with skewness of 0.02, whereas the skewness before masking was 14.47 with a thicker tail at the bright end.

\textcolor{black}{Despite our} rigorous masking procedure, there may remain diffuse light originating from the galaxy outskirts. Indeed, the pixels right outside of the mask \textcolor{black}{appear} to show some remaining light from \textcolor{black}{the} galaxies (see Figure \ref{fig:f_expMask}). The light emanating close to \textcolor{black}{the} galaxies could also be partly composed of ICL\textcolor{black}{;} however\textcolor{black}{,} for a more conservative ICL detection\textcolor{black}{,} we will eliminate \textcolor{black}{all of this light}. We assume that there is no ICL outside the 300 kpc region, and proceed to check how large the mask should be to cover the remaining light from the galaxy outskirts.

Outside of the \textcolor{black}{region of} 300 kpc in the north-east half of the image, we select \textcolor{black}{ten} galaxies including \textcolor{black}{three} member galaxies and \textcolor{black}{seven} SDSS galaxies without spectroscopic redshift and measure the fluxes immediately outside their masked regions. We expand the mask \textcolor{black}{in a} pixel by pixel \textcolor{black}{manner;} i.e., pixels next to (left, right, up and down) \textcolor{black}{the} masked pixel will be masked in the next step. The advantage of th\textcolor{black}{is} mask\textcolor{black}{-}expan\textcolor{black}{sion} method is that we do not assume or model the shape\textcolor{black}{s} of the bright sources but expand the mask as it is in a shape\textcolor{black}{-}independent manner. The images \textcolor{black}{in} Figure \ref{fig:f_expMask} show how the remaining fluxes around an example galaxy are eliminated as the mask is enlarged from 0 pixel\textcolor{black}{s} (original mask) \textcolor{black}{to five} pixels, \textcolor{black}{ten} pixels and finally to 15 pixels. \textcolor{black}{Regarding} the decision of the proper size of the mask which eliminates galaxy outskirt light safely while keeping enough pixels to analyse, we plot the radial profiles of the \textcolor{black}{ten} galaxies \textcolor{black}{while} applying each step of the enlarged mask. Reducing noise while preserving the overall signal tendency, we set the radial bin size \textcolor{black}{to} 8 and take the median value only if the remaining pixels \textcolor{black}{exceed} half of the pixels in the radial bin. We read out the first measured flux value for each mask\textcolor{black}{-}expanding step, i.e., the flux \textcolor{black}{just} outside of that mask, which is plotted \textcolor{black}{with the} gr\textcolor{black}{e}y lines in the bottom panel of Figure \ref{fig:f_expMask}. The fit\textcolor{black}{s} to the \textcolor{black}{remaining fluxes of the ten} galax\textcolor{black}{ies} are plotted \textcolor{black}{with the} red line. As expected, as the mask size grows \textcolor{black}{as} the remaining flux approaches the background value. After 15 steps \textcolor{black}{of} expanded masking\textcolor{black}{,} the fitting value is inside the background error\textcolor{black}{,} which is our detection limit. Thus, we decide to apply the 15 steps \textcolor{black}{of} expanded mask\textcolor{black}{ing in} our analysis. The final mask\textcolor{black}{-}applied postage stamp image in Figure \ref{fig:f_expMask} shows again that the residual light from galaxy outskirt is well removed. Assuming \textcolor{black}{that} the \textcolor{black}{outskirt light sources of the} galaxies in the central region fade out similarly, we apply this oversized mask to the entire image area.

\textcolor{black}{It should be n}ote\textcolor{black}{d} that the background value and \textcolor{black}{corresponding} error are \textcolor{black}{a}ffected by the mask size. We apply the expanded mask to measure the background value (see Section 3.3) and plot the remaining fluxes iteratively, which does not change the tendency. After applying the expanded mask, the pixel brightness distribution becomes even more Gaussian with skewness of 0.005 (see the dark green histogram and the red dashed line in Figure \ref{fig:f3}).

\subsection{Background level and error determination}

\textcolor{black}{R}obust detection of ICL can be claimed only if we measure a significant ICL signal above the background fluctuation level. Thus, in this study the determination of the background level and \textcolor{black}{the} measur\textcolor{black}{ement of} its noise are critical steps. 

The characterization of the background and background noise is performed as follows. The final mask (\textcolor{black}{as} decided \textcolor{black}{upon} in Section 3.2) is applied to each band image. Outside the BCG's 300 kpc region in the north-east half of the image, \textcolor{black}{which we used as a} background in the masking test \textcolor{black}{(see the red defined region in Figure \ref{fig:f1})}, we compute the median value of 30 $\times$ 30 pixels ($\sim$ 28.5 kpc; $\sim4.8\arcsec$) grid boxes. 
The box size is tested \textcolor{black}{while} varying \textcolor{black}{it} from 1 $\times$ 1 pixel to 100 $\times$ 100 pixels, and \textcolor{black}{is} decided considering the apparent galaxy size in the image. Adopting the median values and the \textcolor{black}{corresponding} standard deviation, we find the background
\textcolor{black}{(12055.03 ADU for \textit{i}--band, 5241.76 ADU for \textit{r}--band)} 
and its error (3.3 ADU for \textit{i}--band, 2.23 ADU for \textit{r}--band). In this study, we adopt these 1 $\sigma$ background uncertainties as the detection threshold of the ICL, which are $\mu_{i}^{limit}$(1$\sigma$, $4.8\arcsec \times 4.8\arcsec$) = 28.7 mag/arcsec$^2$ and $\mu_{r}^{limit}$(1$\sigma$, $4.8\arcsec \times 4.8\arcsec$) = 29.1 mag/arcsec$^2$.

We also \textcolor{black}{conduct a} test to compute the median of 30 $\times$ 30 pixels for 10,000 random positions. \textcolor{black}{Because} the random position measurement \textcolor{black}{results in} a slightly smaller background error, we decide to use the non-overlap grid background/background error measurement method \textcolor{black}{to detect the} ICL \textcolor{black}{more conservatively}.  Furthermore, we split the background region and check if the background level changes depending on the distance to the BCG \textcolor{black}{(see Table \ref{tab:bg_table})}, and subsequently how this affects the surface brightness and colour profiles. We could not detect any trend on the background level related to the distance to the BCG. The different background levels for each radial bin lead to some changes of the ICL profile outside 155 kpc, but for the colour profile, they do not significantly affect the blue colour nor the steep profile (see Section 4.1).

\begin{table}
 \caption{\textcolor{black}{Background level and error in different regions}}
 \label{tab:bg_table}
 \begin{tabular*}{\columnwidth}{@{}c@{\hspace*{8pt}}c@{\hspace*{8pt}}c@{\hspace*{8pt}}c@{\hspace*{8pt}}c@{}}
  \hline
   &  \multicolumn{2}{c}{\textcolor{black}{\textit{i}--band}} & \multicolumn{2}{c}{\textcolor{black}{\textit{r}--band}}\\
   \cline{2-5}
   \textcolor{black}{Background} & \textcolor{black}{Background} & \textcolor{black}{Background}  & \textcolor{black}{Background}  & \textcolor{black}{Background}  \\
  \textcolor{black}{region} & \textcolor{black}{level} & \textcolor{black}{error}&  \textcolor{black}{level} & \textcolor{black}{error}\\
\textcolor{black}{(kpc from centre)}& \textcolor{black}{[ADU]} & \textcolor{black}{[ADU]} &  \textcolor{black}{[ADU]} & \textcolor{black}{[ADU]}\\
  \hline
  \textcolor{black}{All (300 $\sim$)$^c$}& \textcolor{black}{12055.03}& \textcolor{black}{ 3.30 }& \textcolor{black}{ 5241.76 }& \textcolor{black}{ 2.23}\\
  \textcolor{black}{300 $\sim$ 400}& \textcolor{black}{ 12052.63}& \textcolor{black}{ 3.24 }& \textcolor{black}{5241.06 }& \textcolor{black}{1.97}\\
  \textcolor{black}{400 $\sim$ 500}& \textcolor{black}{ 12054.67}& \textcolor{black}{ 2.96 }& \textcolor{black}{5242.25 }& \textcolor{black}{1.27}\\
  \textcolor{black}{500 $\sim$ 600}& \textcolor{black}{ 12056.92}& \textcolor{black}{ 3.17 }& \textcolor{black}{5242.74 }& \textcolor{black}{1.79} \\
  \textcolor{black}{600 $\sim$ 700}& \textcolor{black}{ 12054.42}& \textcolor{black}{ 3.41 }& \textcolor{black}{5240.53 }& \textcolor{black}{2.66}\\
  \hline
  \multicolumn{5}{l}{\textcolor{black}{$^c$ The All (300 $\sim$) corresponds to the red defined region in Figure \ref{fig:f1}.}}\\
 \end{tabular*}
\end{table}

We also estimated the surface brightness limit as suggested by \cite{2020A&A...644A..42R}, which results in a deeper limit \textcolor{black}{of}  $\mu_{i}^{limit}$(1$\sigma$, $4.8\arcsec \times 4.8\arcsec$) = 30.19 mag/arcsec$^2$. The Gaussian standard deviation on a desired angular scale box is \textcolor{black}{in this case} converted from the pixel variation, assuming \textcolor{black}{that} the flux noise follows a normal distribution. In our case, the pixel variation (i.e., \textcolor{black}{a} 1 $\times$ 1 pixel grid box variation in the test above) is 1$\sigma_{pix}$ = 25.01 ADU (see the red dashed line in Figure \ref{fig:f3}).
\textcolor{black}{It should also be} note\textcolor{black}{d} that our standard deviation is directly measured from \textcolor{black}{boxes} 30 $\times$ 30 pixels \textcolor{black}{in size in} the final reduced and masked image. For more conservative ICL measurements, we decide to use our method. 

\section{Results} \label{sec:results}
Below we describe the ICL measurements in Section 4.1, a\textcolor{black}{fter which we} derive \textcolor{black}{the} properties of the galaxy cluster, the BCG and member galaxies thought to be related to the characteristics of the ICL in Section\textcolor{black}{s} 4.2, 4.3 and 4.4, respectively.


\subsection{ICL \textcolor{black}{m}easurements: surface brightness, colour, spatial distribution, fraction}
\begin{figure*}
\centering
\includegraphics[angle=90,trim={4.0cm 0 5.0cm 0},clip,width=2.0\columnwidth]{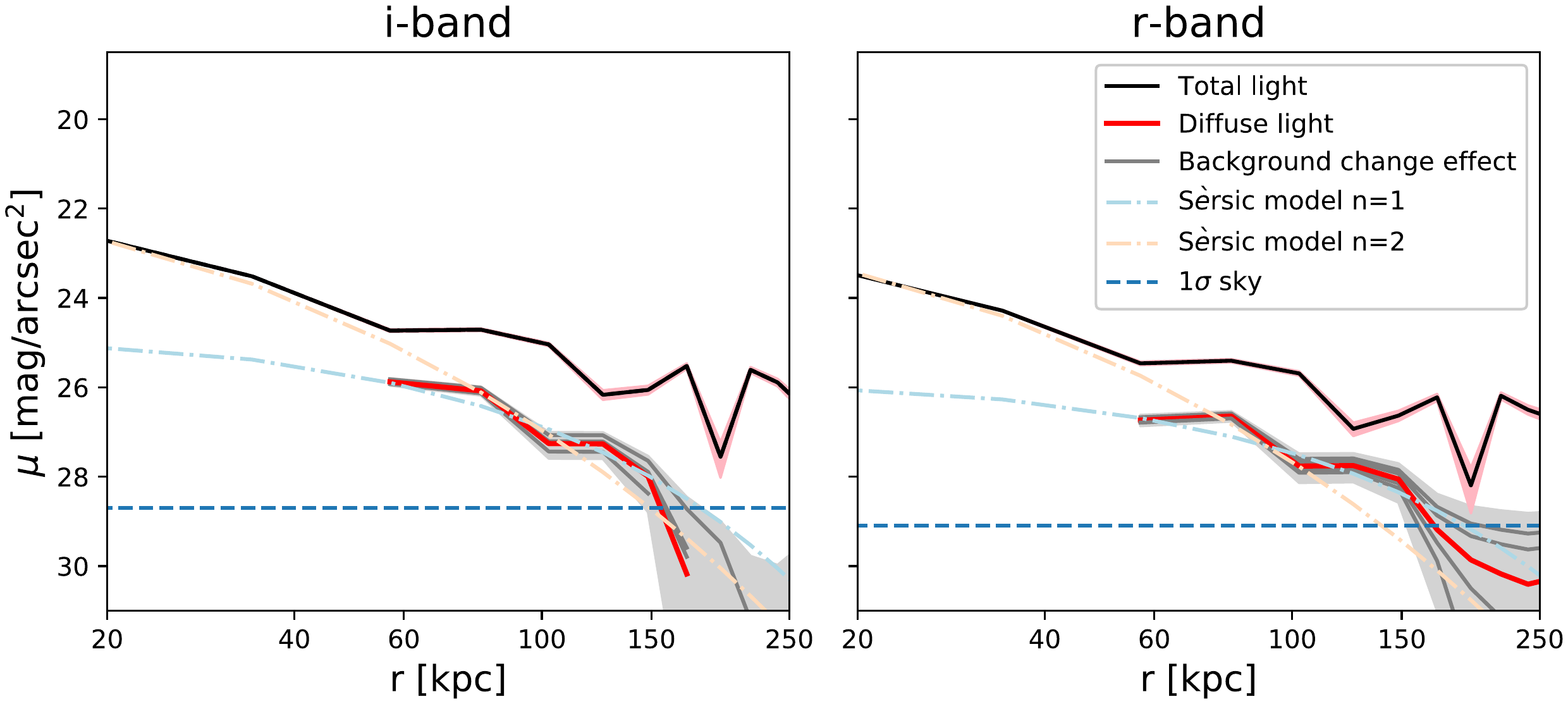}
\caption{\textcolor{black}{R}adial surface brightness profiles for \textcolor{black}{the} \textit{i}--band (\textcolor{black}{l}eft) and \textit{r}--band (\textcolor{black}{r}ight). The mean value of each radial bin for total light (masking non-member galaxies and stars, and expanding the mask) and diffuse light (masking all bright objects, and expanding the mask) are plotted as black and red solid line\textcolor{black}{s}, respectively (see Section\textcolor{black}{s} 3.2 and 4.1 for detail\textcolor{black}{s}). \textcolor{black}{For the \textit{i}--band, the profile in the ranges, where the averaged fluxes have negative values with large uncertainties is not plotted.} The s$\grave{e}$rsic models with s$\grave{e}$rsic index\textcolor{black}{es of} n = 1 and 2 are plotted as \textcolor{black}{the} light blue and orange dot-dashed line, respectively. The 1$\sigma$ level detection limit is \textcolor{black}{indicated by the} blue dashed line, \textcolor{black}{and} the errors \textcolor{black}{for total (diffuse) light} are plotted as \textcolor{black}{the} filled \textcolor{black}{pink (grey)} region\textcolor{black}{s}. \textcolor{black}{The effects of the background level changes (see the Table \ref{tab:bg_table}) on the diffuse light profiles are plotted as grey solid lines.} The \textcolor{black}{analysed} radius range of the diffuse light starts from 60 kpc and ends by \textcolor{black}{155 kpc, where the \textit{i}--band surface brightness profile touches the detection limit.} 
\label{fig:f5}}
\end{figure*}

In nearby clusters, ICL is studied typically below a surface brightness threshold of $\mu_V\sim$ 26.5 mag/arcsec$^2$ in the rest-frame \citep{2005ApJ...631L..41M, 2017ApJ...834...16M, 2011ApJ...732...48R}. This limit corresponds to $\mu_{i}\sim$ 27.18  mag/arcsec$^2$ and  $\mu_{r}\sim$ 28.08 mag/arcsec$^2$ at $z= 0.47$ considering the \textcolor{black}{surface brightness} dimming \textcolor{black}{correction of} $\sim \textcolor{black}{2.5} (1+z)^{\textcolor{black}{3}}$ and taking into account the stellar population evolution and the passband shift at the cluster redshift. \textcolor{black}{These surface brightness limit conversions are calculated following equation (1) and (2) in \cite{2015MNRAS.449.2353B}.} Here\textcolor{black}{,} we used the stellar population synthesis models of \cite{2003MNRAS.344.1000B} while assuming the formation redshift of $z_f$ = 3, a simple stellar population with solar metallicity, and the initial mass function of \cite{2003PASP..115..763C}.
Thus, with our surface brightness limits ($\mu_{i}$ = 28.7 mag/arcsec$^2$ and $\mu_{r}$ = 29.1 mag/arcsec$^2$), our Gemini images allow us to explore the ICL distribution \textcolor{black}{out} to $\sim$ 155 kpc (170 kpc) from the BCG for \textcolor{black}{the} \textit{i}--band (\textit{r}--band)\textcolor{black}{,} as shown in Figure \ref{fig:f5}.

\textit{\bf Surface brightness}: In Figure \ref{fig:f5}, we plot \textcolor{black}{the} radial surface brightness profiles for the ICL as a function of \textcolor{black}{the} radius from the BCG. 
The radial profile is derived by taking the azimuthally averaged value of \textcolor{black}{the} remaining pixels in each bin after applying the mask. The inner radius was determined by locating the bin \textcolor{black}{in which} unmasked pixels start to dominate\textcolor{black}{. Accordingly}, the ICL profile does not have values in the central region of 60 kpc. The outer radius corresponds to the detection limit, 28.7 mag/arcsec$^2$, around 155 kpc in \textcolor{black}{the} \textit{i}--band case. Therefore, the ICL analysis is conducted only \textcolor{black}{in the range of} 60 kpc and 155 kpc for both \textit{i}-- and \textit{r}--band images.

We also plot \textcolor{black}{the} radial surface brightness profiles for the total cluster light together after masking out stars and non-member galaxies. We identify the spectroscopically confirmed member galaxies (see Section 4.2.1)\textcolor{black}{, after which} the mask is expanded in the same way as the ICL. There \textcolor{black}{may be} more undetected member galaxies, \textcolor{black}{possibly causing us to} underestimat\textcolor{black}{e} the total light. \textcolor{black}{In fact}, around 125 kpc \textcolor{black}{and 195 kpc} there is no member galaxy in the bin, which results in a drop \textcolor{black}{of} the total light. 
\textcolor{black}{In contrast}, the total light \textcolor{black}{showed} an excess \textcolor{black}{level} around 170 kpc due to several member galaxies in the bin.

The BCG-dominated central part of the total light is fitted as \textcolor{black}{an} s$\grave{e}$rsic profile with \textcolor{black}{an} s$\grave{e}$rsic index of n = 2 (further described in Section 4.3), whereas the ICL is better described by the s$\grave{e}$rsic model with n = 1 (Figure \ref{fig:f5}). This result suggests that the two components (BCG+ICL) have different physical origin\textcolor{black}{s}  \citep[][]{2015MNRAS.451.2703C}, where the outer component consists mainly of the ICL rather than the BCG.
To disentangle the ICL from the BCG outskirt directly (i.e., to check if it follows the galaxy cluster potential or not), we need its spectroscopic data \textcolor{black}{to} measure the velocity dispersion, which was not possible in this study.

\begin{figure}
\centering
\includegraphics[trim={0.0cm 8cm 0.0cm 9cm},clip,width=1.0\columnwidth]{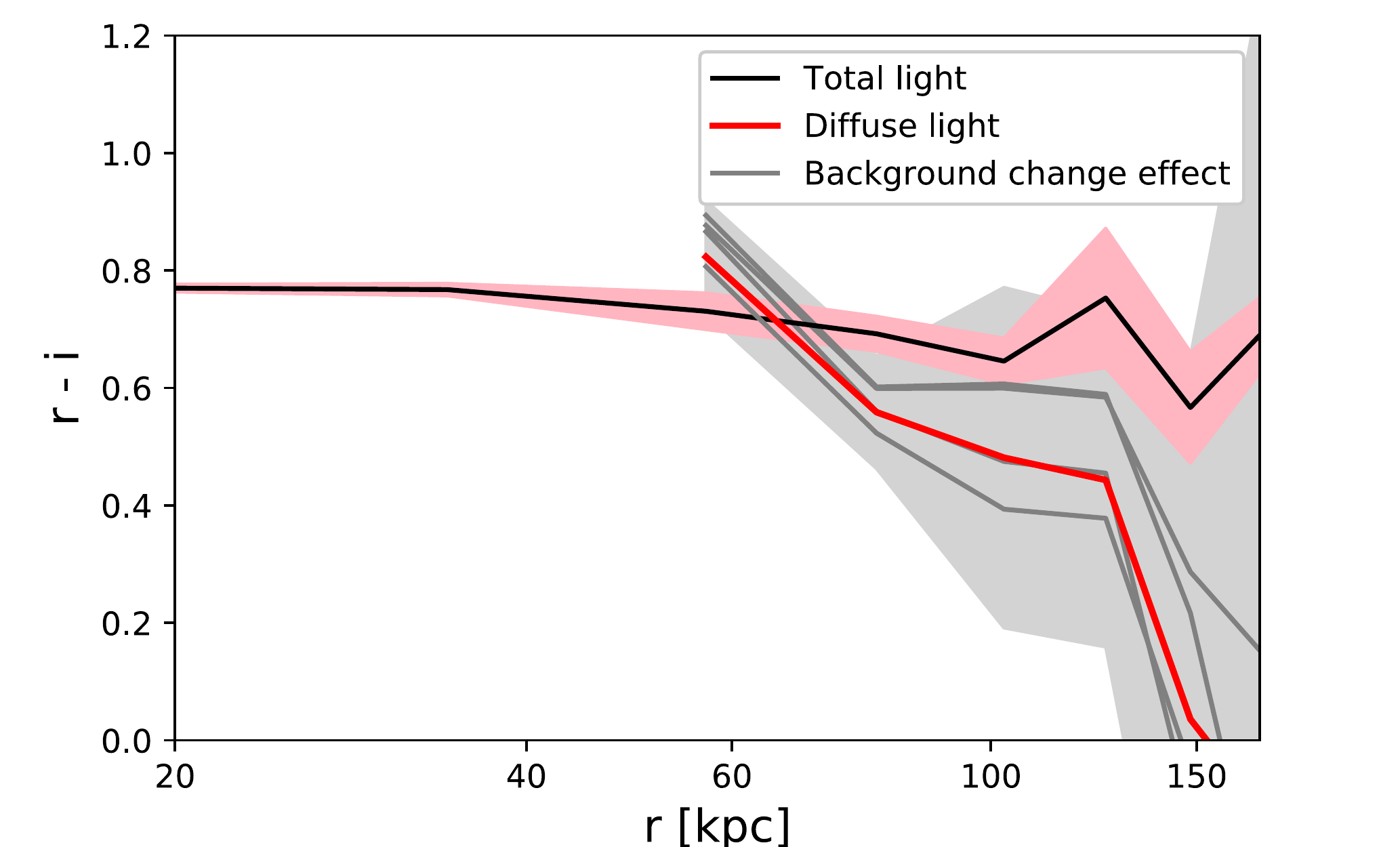}
\caption{The \textit{r}--\textit{i} colour profile for \textcolor{black}{the} total light (black solid line) and diffuse light (red solid line). The definition of the total/diffuse light, the radial bin width\textcolor{black}{,} and the radius range are \textcolor{black}{identical to those in} Figure \ref{fig:f5}. The \textcolor{black}{pink (grey)} shaded regions show \textcolor{black}{the Monte-Carlo} errors of the \textit{r}--\textit{i} colour \textcolor{black}{of the total (diffuse) light, based on the background uncertainty of each band.} \textcolor{black}{The effects of the background level changes (see the Table \ref{tab:bg_table}) on the colour profile of the diffuse light are plotted as grey solid lines.} U\textcolor{black}{p to} 70 kpc\textcolor{black}{,} the colour of the diffuse light is similar to \textcolor{black}{that of} the total light, whereas outside of 80 kpc it becomes bluer than \textcolor{black}{the} total light\textcolor{black}{,} including the BCG and other nearby red member galaxies. The colour gradient of the diffuse light is steeper than that of the total light as well\textcolor{black}{, but the errors are large and preclude definitive results}.  
\label{fig:f6}}
\end{figure}

\textit{\bf Colour}: After \textcolor{black}{identical} masking, we created \textit{r}--\textit{i} colour profiles \textcolor{black}{of} the total light and ICL by subtracting the surface brightness profiles at the same radial bin in two bands (Figure \ref{fig:f6}), which corresponds to \textcolor{black}{the} \textit{B}--\textit{V} colour in the rest-frame. 
\textcolor{black}{Regarding} the error of the colour profile, we used the Monte-Carlo method based on the background estimates and uncertainty. As the Monte-Carlo estimated error of the colour profile becomes significantly larger in the outer radius, we do not trust the colour \textcolor{black}{greater than} $\sim$ 150 kpc. \textcolor{black}{The choice of the background level (see the Table \ref{tab:bg_table}) varies the colour maximally around $\sim$ 0.2 in the range of 80 $\sim$ 130 kpc (see the grey solid lines in Figure \ref{fig:f6}), which does not significantly affect our results.}

The colour of the diffuse light around 60 $\sim$ 80 kpc is similar to those of \textcolor{black}{the} BCG and bright red members. This might imply that diffuse light in the central region of the cluster is closely tied to the BCG and/or bright red members. The total light gradually becomes bluer beyond $\sim$ 70 kpc, but this trend \textcolor{black}{appears to be} steeper for the diffuse light. In the range of 80 $\sim$ 130 kpc\textcolor{black}{,} the ICL colour is \textit{r}--\textit{i} = \textcolor{black}{0.47 $\pm$ 0.28}, whereas the total light colour is \textit{r}--\textit{i} = \textcolor{black}{0.68 $\pm$ 0.05}. This difference might suggest that the main stellar population of the diffuse light is not likely to originate from that of the BCG and/or bright red members. In Section 5.1 we discuss the possible origins of the ICL \textcolor{black}{further}, specifically the relation between the colour of the ICL and that of the BCG/member galaxies and the colour evolution of the ICL.

For the total light, the spectroscopically confirmed member galaxies could be selectively red galaxies, \textcolor{black}{as} they are more likely to be brighter (more massive) red-sequence member galaxies \textcolor{black}{and} thus listed \textcolor{black}{as a} priority for many spect\textcolor{black}{r}oscopic observations. This could bias the colour of the total light toward red. If we recalculate \textcolor{black}{while} including all detected galaxies, the colour of total light \textcolor{black}{becomes} \textit{r}--\textit{i} = \textcolor{black}{0.65 $\pm$ 0.04}, which is still redder than the colour of the ICL.

\begin{figure*}
\includegraphics[height=2.45in,trim={0.4cm 0 0.7cm 0},clip]{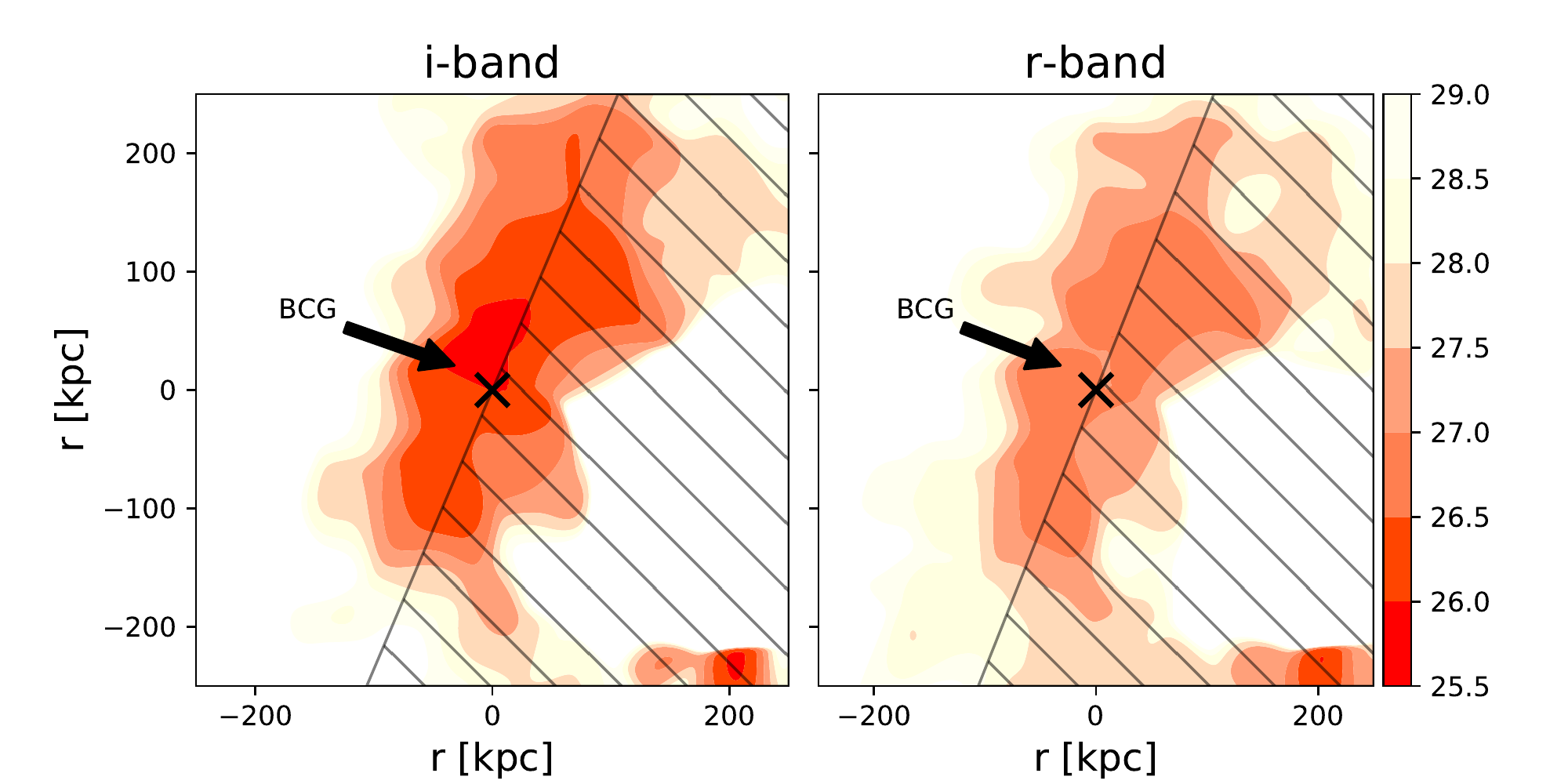}
\includegraphics[height=2.45in,trim={1.2cm 0 0 0},clip]{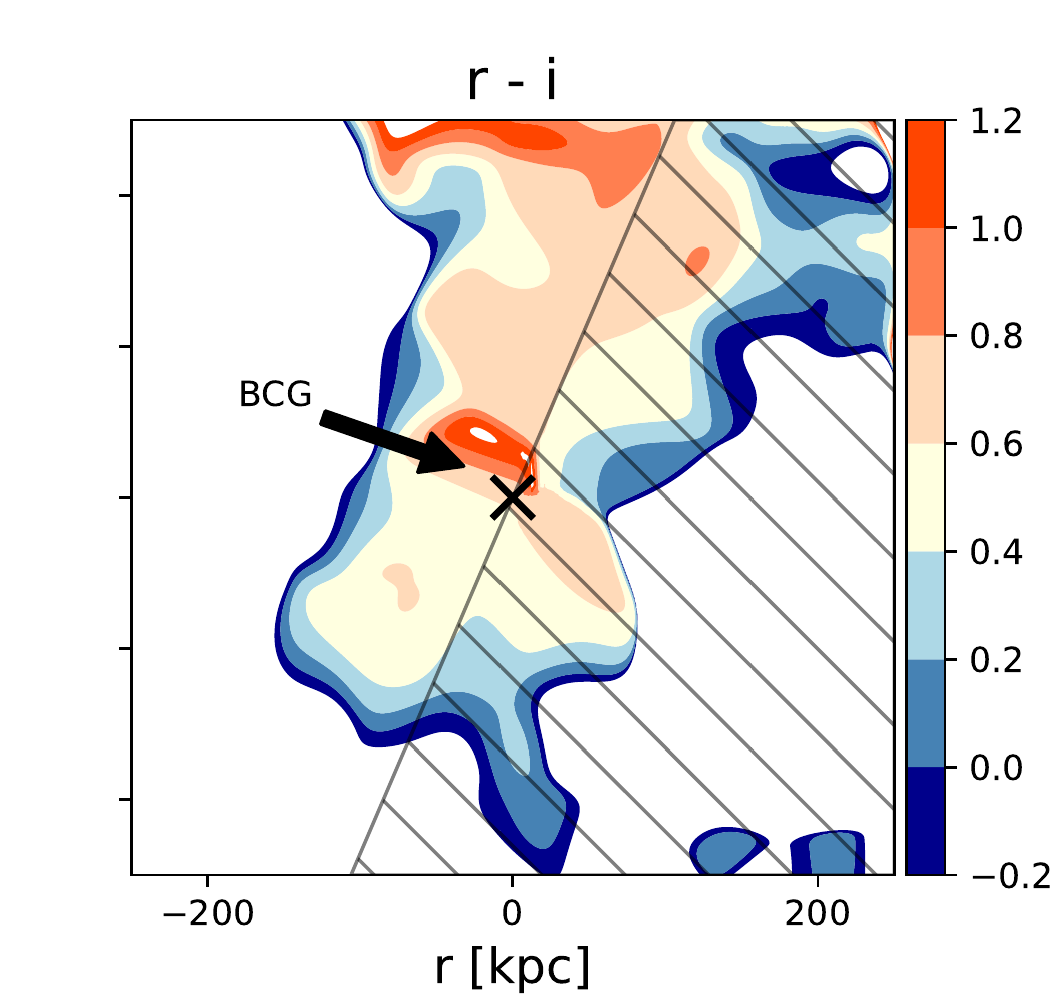}
\caption{\textcolor{black}{D}iffuse light (after masking) surface brightness 2D map for \textcolor{black}{the} \textit{i}--band (\textcolor{black}{l}eft) and \textit{r}--band (\textcolor{black}{m}iddle) and \textcolor{black}{the} \textit{r}--\textit{i} colour 2D map (\textcolor{black}{r}ight) are plotted. The colour bar for \textcolor{black}{the} \textit{i}--band and \textit{r}--band 2D maps \textcolor{black}{is} locate\textcolor{black}{d} on the right side of \textcolor{black}{the} \textit{r}--band 2D map, which has unit of mag/arcsec$^2$. The colour bar of \textit{r}--\textit{i} colour 2D map is on the right side of \textcolor{black}{this map}. The 2D maps are Gaussian smoothed using the \texttt{Gaussian2DKernel astropy package} with an oversampling factor of 17.
The BCG centre location is marked. The hatched region on the south-west side is excluded during the ICL analysis.
Both the \textit{i}--band and the \textit{r}--band diffuse light distribution\textcolor{black}{s} show clear elongation and \textcolor{black}{a} tilted position angle. \label{fig:f7}}
\end{figure*}

\textit{\bf Spatial distribution}: The spatial distribution of the diffuse light is shown in Figure \ref{fig:f7}. Similar to the colour radial profile, subtracting the \textit{i}--band surface brightness from the \textit{r}--band gives us \textcolor{black}{a two}-dimensional colour map of the diffuse light. Th\textcolor{black}{e}se surface brightness and colour 2D maps are smoothed by \textcolor{black}{the} \texttt{Gaussian2DKernel astropy package} with an oversampling factor of 17. \textcolor{black}{Due to} the masking and smoothing, \textcolor{black}{some} small part\textcolor{black}{s} of the central region in \textcolor{black}{the} 2D maps are not plotted. Both the \textit{i}--band and \textit{r}--band diffuse light distributions show \textcolor{black}{a} distinct elongated and tilted structure. This structure is visible in the \textit{r}--\textit{i} colour distribution as well. 
We compare it with the distribution of the member galaxies/luminosity\textcolor{black}{-}weighted distribution of member galaxies in Section 5.2.

\begin{figure}
\centering
\includegraphics[trim={0.0cm 8cm 0.0cm 8cm},clip,width=1.0\columnwidth]{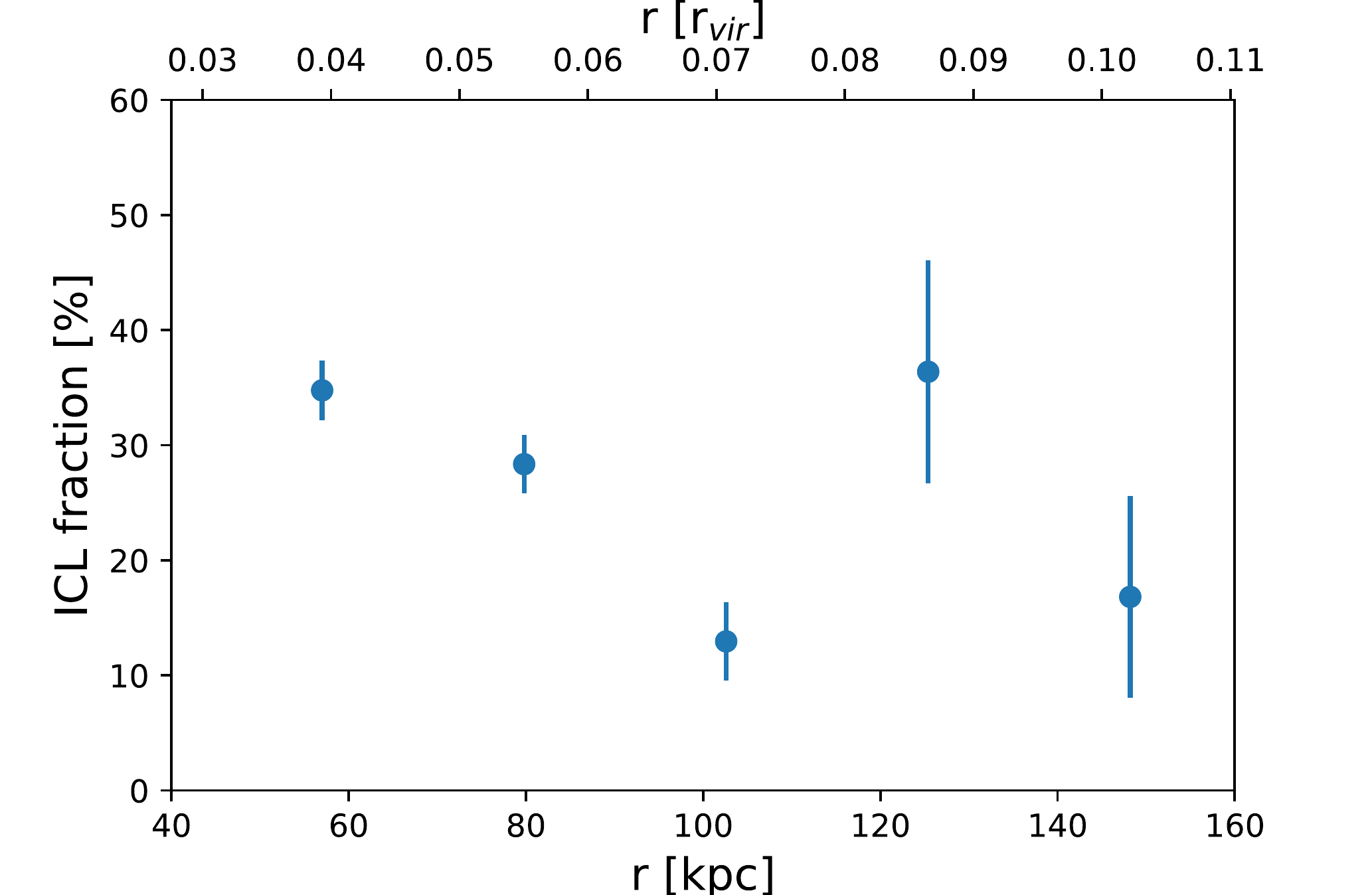}
\caption{Radial ICL fraction calculated \textcolor{black}{according to} the ratio \textcolor{black}{of the} diffuse light \textcolor{black}{to the} total light. \textcolor{black}{The r}adius in \textcolor{black}{unit of the} virial radius is shown in \textcolor{black}{the} upper axis. The background\textcolor{black}{-}variation\textcolor{black}{-}driven ICL fraction errors are marked as error bars. Through mask incompleteness correction (described in Section 3.2)\textcolor{black}{,} we eliminated \textcolor{black}{the} possible remaining light from \textcolor{black}{the} galaxy outskirt\textcolor{black}{s}. \label{fig:f9}}
\end{figure}

\textit{\bf ICL fraction}: In Figure \ref{fig:f9}, we show the ICL fraction in \textcolor{black}{the} \textit{i}--band (corresponds to \textcolor{black}{the} \textit{V}--band in the rest-frame)\textcolor{black}{,} defined as the ratio of the surface brightness of the ICL to the total light at the same radial bin. In order to measure the total cluster light, we mask out non-member galaxies and stars\textcolor{black}{,} as described \textcolor{black}{earlier}. Again, the lack of spectroscopically confirmed member galaxies in the bin around 125 kpc make\textcolor{black}{s} the total light lower and \textcolor{black}{enhances} the ICL fraction. 

\textcolor{black}{By m}aking the bin range larger, in the range of 60 $\sim$ 155 kpc, we calculate \textcolor{black}{an} ICL fraction of J1054 is $15.07 \pm 4.57 \%$. The same calculation for \textcolor{black}{the} \textit{r}--band gives \textcolor{black}{an} ICL fraction of $19.95 \pm 5.67 \%$. We could interpret this as a conservative estimation of \textcolor{black}{the} ICL fraction due to our rigorous expanding mask procedure to cover diffuse light that may belong to galaxies. However, the possible missing member galaxy light in this area would lead to an underestimation of the total light. Similarly\textcolor{black}{,} for the colour estimation, if we calculate the ICL fraction including all detected galaxies for the total light, the ICL fraction becomes $14.50 \pm 4.4 \%$  for \textcolor{black}{the} \textit{i}--band, and $18.86 \pm 5.36 \%$ for \textcolor{black}{the} \textit{r}--band, which could correspond to a lower limit of the ICL fraction for J1054.

We exclude the range of 0 $\sim$ 60 kpc when calculating the overall ICL fraction of J1054, where the abundance of ICL could not be estimated. In this range, the BCG light should be dominant, but at the same time, the ICL abundance is also expected to \textcolor{black}{reach its} maximum along the line of sight. \textcolor{black}{Because} separation between \textcolor{black}{the} ICL and the extended envelope of the BCG is photometrically impossible  \citep[as discussed in][]{2007ApJ...666..147G}, to mitigate the uncertainty, we focus on the region where the ICL is more dominant. However, for comparison\textcolor{black}{s} with other studies, we calculate the ICL fraction using \textcolor{black}{a} surface brightness cut in the range 0 < r < $R_{500}$ (in Section 5.3), which results in a rather high ICL fraction of J1054.

\subsection{Galaxy cluster properties}
\subsubsection{Member identification}
\begin{figure}
\centering
\includegraphics[trim={0.0cm 9cm 0.0cm 9cm},clip,width=1.0\columnwidth]{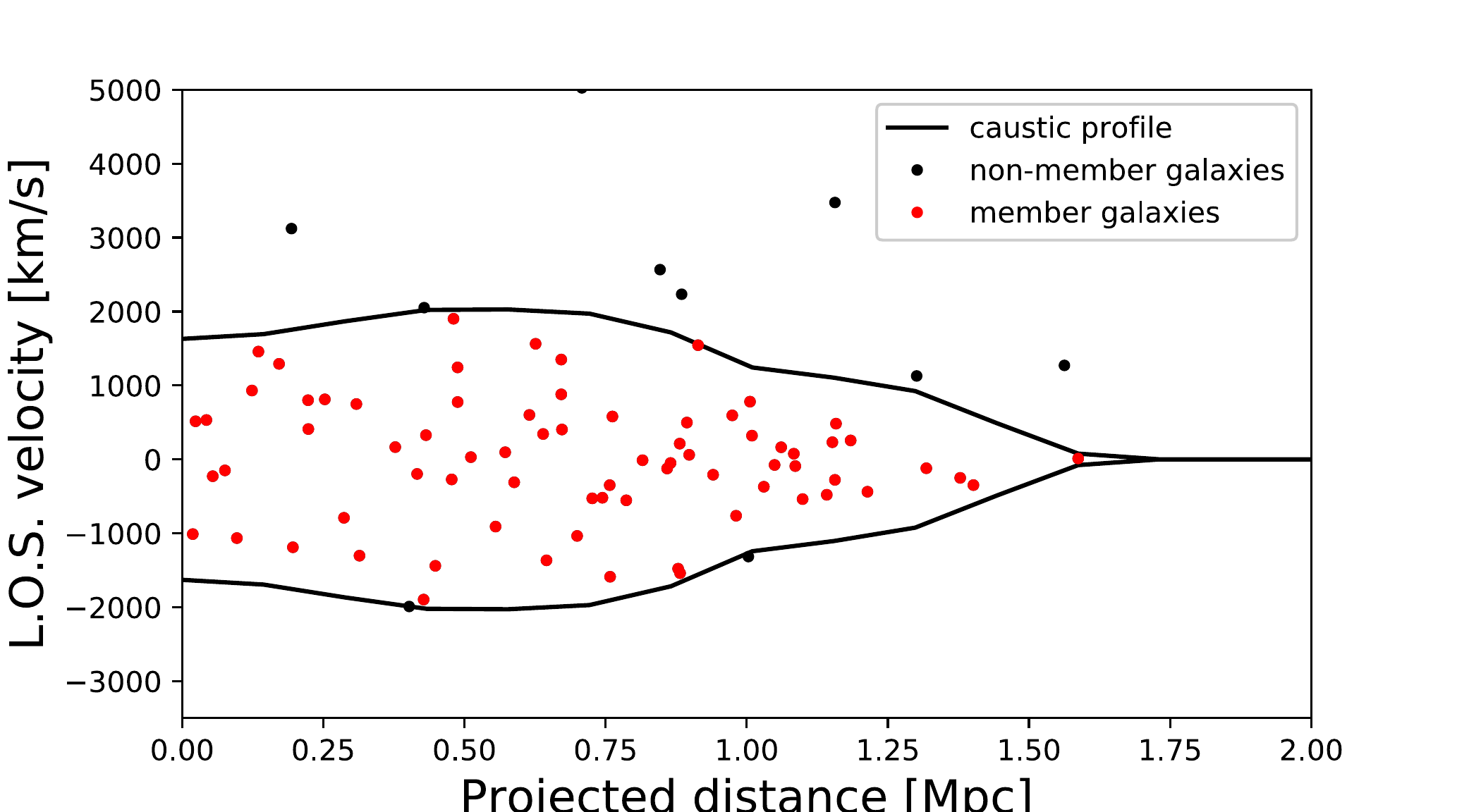}
\caption{\textcolor{black}{P}hase space diagram of J1054. Among 119 candidate galaxies with spectroscopic redshift\textcolor{black}{s}, 73 member galaxies (red dots) are selected through \textcolor{black}{the} Caustic method. The calculated caustic profile from \textcolor{black}{the velocity information of these} 119 \textcolor{black}{galaxies} is plotted as \textcolor{black}{the} black solid line \textcolor{black}{in the figure}. \textcolor{black}{N}on-member galaxies with line-of-sight velocit\textcolor{black}{ies of less} than -3500 km/s or \textcolor{black}{that exceed} 5000 km/s are not plotted in this diagram. \label{fig:f4}}
\end{figure}

The previous study of J1054 obtained 116 spectroscopic redshifts of galaxies\textcolor{black}{,} and of those 78 member galaxies were identified \citep{2011A&A...527A.143A}. Aguerri et al. used the shifting gapper method, which rejects galaxies with large peculiar velocities in fixed distance bins. We add \textcolor{black}{three} more redshifts from our MOS observation to the previous 116 and then re-select member galaxies using the Caustic method, which is a physically well\textcolor{black}{-}motivated technique \citep{2014Caustic}. The Caustic method measures the escape velocity profile\textcolor{black}{s} of galaxy clusters and identifies member galaxies based on \textcolor{black}{these profiles}. In the Caustic method, we choose the Antonaldo Diaferio's threshold \citep{1999MNRAS.309..610D} for cluster candidate members and fix\textcolor{black}{ed} the centre as \textcolor{black}{the} BCG location, result\textcolor{black}{ing} in 73 newly selected member galaxies. Compare\textcolor{black}{d} to the previous result with 78 members, our membership selection \textcolor{black}{process} is tighter \textcolor{black}{and} results \textcolor{black}{in fewer} member galaxies. Figure \ref{fig:f4} shows \textcolor{black}{a} phase space diagram plotting the 119 galaxies with \textcolor{black}{a} spectroscopic redshift (line-of-sight velocit\textcolor{black}{ies lower} than -3500 km/s or \textcolor{black}{higher} than 5000 km/s are not plotted) \textcolor{black}{as well as the corresponding} escape\textcolor{black}{-}velocity\textcolor{black}{-}driven caustic profile\textcolor{black}{s} and selected member galaxies inside \textcolor{black}{the} caustic profile. The line-of-sight velocity of the selected member galaxies \textcolor{black}{ranges} from -1896 km/s to 1902 km/s, and the maximum projected distance from \textcolor{black}{the corresponding} BCG centre is 1.59 Mpc.      

\subsubsection{Cluster mass and radius derivation}
Using the velocity data of \textcolor{black}{the} identified member galaxies, the Caustic method provides a median redshift of $ z  = 0.465729$ \textcolor{black}{$\pm$ 0.003995}, \textcolor{black}{a} median line-of-sight velocity of $ v  = 139622.0 $ \textcolor{black}{$\pm$ 817.13} km/s, \textcolor{black}{a} rest-frame velocity dispersion $\sigma_v = 934.36$  \textcolor{black}{$\pm$ 13.20} km/s and $r_{200} = 1.68$  \textcolor{black}{$\pm$ 0.20} Mpc, $M_{200}= \textcolor{black}{7.58 \pm 1.40}\times 10^{14} M_\odot$ from fitting \textcolor{black}{with} an NFW profile.
Those results are not far from those \textcolor{black}{in a} previous study \citep{2011A&A...527A.143A} with 78 member galaxies. Again, this confirms that J1054 is an extremely rare target\textcolor{black}{,} being a fossil and massive cluster at \textcolor{black}{an} intermediate redshift \textcolor{black}{level}.
\subsection{BCG properties}
\begin{figure*}
\centering
\includegraphics[height=2.45in,trim={2.0cm 0 1.0cm 0},clip]{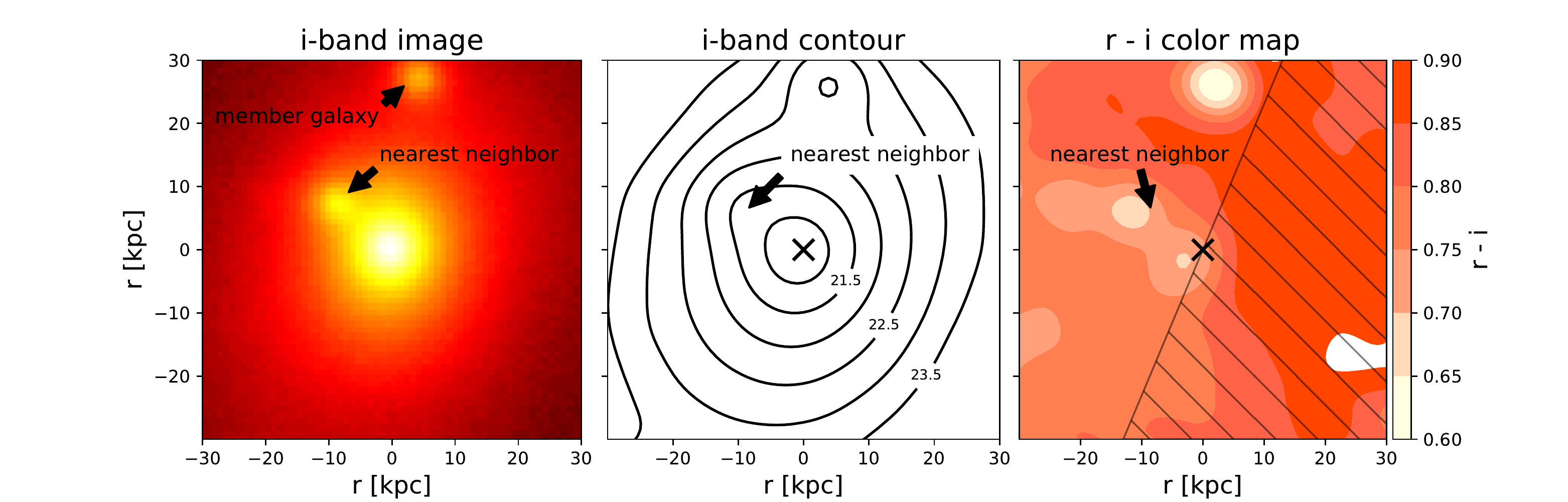}
\caption{Stamp image (\textcolor{black}{l}eft), surface brightness contour in \textcolor{black}{the} \textit{i}--band (\textcolor{black}{m}iddle) and \textit{r}--\textit{i} colour contour (\textcolor{black}{r}ight) of the BCG. The colour bar for \textcolor{black}{the} \textit{r}--\textit{i} colour \textcolor{black}{is} locate\textcolor{black}{d} on the right side of \textcolor{black}{the contour map}. The hatched region on the south-west side is excluded for the ICL analysis. The nearest neighbo\textcolor{black}{u}r of the BCG is detected at \textcolor{black}{approximately}10 kpc from the BCG centre. No spectroscopic redshift information exist\textcolor{black}{s} for this structure, but it shows \textcolor{black}{an} \textit{r}--\textit{i} colour \textcolor{black}{similar to that of} the centre of the BCG. \label{fig:f11}}
\end{figure*}

Our deep imaging observations of J1054's BCG reveal a distinct structure that resembles a nearest neighbo\textcolor{black}{u}r (or accretion) \textcolor{black}{approximately} 10 kpc away from the BCG centre (see Figure \ref{fig:f11}). Its colour is \textcolor{black}{identical to that of} the BCG centre, which suggests that it is likely to be\textcolor{black}{long to} the cluster. Unfortunately, our MOS slit on the BCG does not go through this structure, \textcolor{black}{meaning that} we could not analy\textcolor{black}{s}e its spectrum. Future spectroscopic stud\textcolor{black}{ies} on this structure would help to reveal \textcolor{black}{its} identity and \textcolor{black}{the} dynamic relation with the BCG.
The existence of this structure \textcolor{black}{wa}s not reported in \textcolor{black}{a} previous study of J1054 \citep{2011A&A...527A.143A}, though they report that the BCG does not have extra light outside and can be fitted as \textcolor{black}{a} single s$\grave{e}$rsic profile with \textcolor{black}{an} index \textcolor{black}{of} n $\sim$ 2.1, which means it is not a cD galaxy. They conclude that to have this low s$\grave{e}$rsic index as a luminous elliptical galaxy at this redshift, it should have undergone extreme wet merging ($\sim$ 80\% gas rich). Using the \texttt{Sersic1D astropy package}, we measure the s$\grave{e}$rsic index of the observed BCG image with and without masking \textcolor{black}{of} the neighbo\textcolor{black}{u}r structure. We do not fit in the inner-most seeing-dominated region  \citep{1999ApJ...516L..61M, 2017A&A...603A..38S} of the BCG inside \textcolor{black}{nine} kpc ($\sim$ 1.5 $\times$ FWHM). The result\textcolor{black}{ing} s$\grave{e}$rsic index is n \textcolor{black}{= 2.21 $\pm$ 0.02} with the neighbo\textcolor{black}{u}r structure and n \textcolor{black}{= 2.00 $\pm$ 0.02} without it, \textcolor{black}{showing a minor} difference. 

\begin{figure*}
\centering
\includegraphics[angle=90,height=2.45in,trim={5.5cm 0 5.5cm 0},clip]{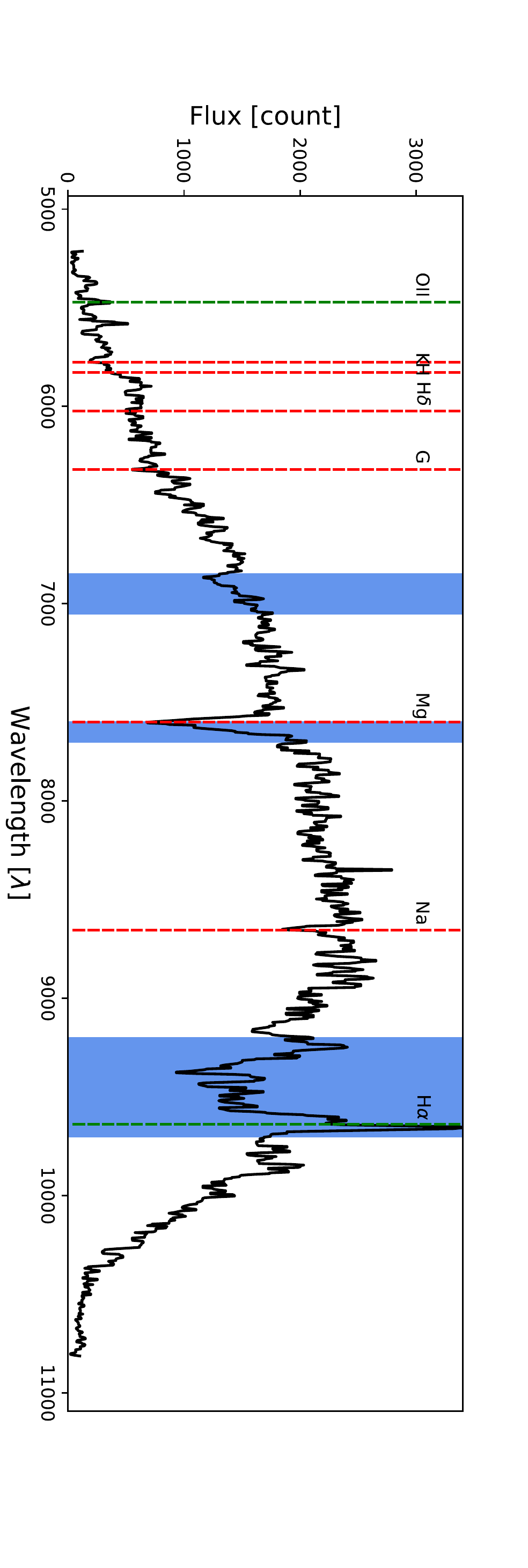}
\caption{\textcolor{black}{The} BCG spectrum from the MOS observation is plotted. The green and red dashed lines are \textcolor{black}{the} emission and absorption lines for $z=0.468$, respectively. The blue filled regions are where the sky lines are located. Mg \textcolor{black}{and} Na absorption lines and \textcolor{black}{a} weak [OII] emission line are visible, but \textcolor{black}{the} H$\delta$ absorption line is negligible. \label{fig:f12}}
\end{figure*}

We analy\textcolor{black}{s}e the BCG spectrum from the MOS observation (see Figure \ref{fig:f12}). Using the Mg \textcolor{black}{and} Na absorption lines and the weak [OII] emission line\textcolor{black}{,} we confirm its redshift as 0.468. The [OII] emission line is very weak and there is no visible H$\delta$ absorption line, which implies that there was no post-starburst activity within 1 Gyr. The strong H$\alpha$ emission line that appeared similar\textcolor{black}{ly} in all observed galaxies is less reliable \textcolor{black}{because} it is located where sky lines \textcolor{black}{exist}.
The \textit{r}--\textit{i} colour of \textcolor{black}{the} BCG is 0.89 \textcolor{black}{$\pm$ 0.04} (\texttt{MAG\_AUTO} of \texttt{SExtractor} from \textcolor{black}{the} observed Gemini image), which is comparable to \textcolor{black}{those of} other observed luminous red elliptical galaxies at $ z\sim 0.47$ \citep[\textit{r}--\textit{i} = 0.8/0.9 at $z\sim 0.45/0.5$;][]{2009MNRAS.394L.107M}.
We calculate its $D_n 4000$ index following the definition of \cite{1999ApJ...527...54B}, which is the flux within a 100 \AA{}  window with \textcolor{black}{a} central wavelength of 4050 \AA{}  divided by the flux in a 100 \AA{}  window with \textcolor{black}{a} central wavelength of 3900 \AA.
The $D_n 4000$ index of the BCG is 1.98 from our Gemini spectrum and 1.71 from \textcolor{black}{the} SDSS spectrum, which is higher or comparable to \textcolor{black}{those of} other passive luminous red galaxies with $D_n 4000$ =1.75 \citep{2006MNRAS.373..349R} \textcolor{black}{and} mean\textcolor{black}{ing} that the BCG of J1054 is composed of an older stellar population. 
\textcolor{black}{We fitted the spectrum using \texttt{CIGALE} \citep{2005MNRAS.360.1413B, 2009A&A...507.1793N, 2019A&A...622A.103B}, which estimates the BCG age to be 7 $\pm$ 0.5 Gyr.}

We also check \textcolor{black}{data from} the Wide-field Infrared Survey Explorer \citep[WISE;][]{2010AJ....140.1868W}, especially \textcolor{black}{the} \texttt{w1mag} [3.4$\mu$m] - \texttt{w3mag} [12$\mu$m] colour of the BCG, \textcolor{black}{as} the mid-infrared (mid-IR) colour\textcolor{black}{s} of red early-type galaxies have strong correlations in age-sensitive spectral features\textcolor{black}{, meaning that} the mid-IR can be a useful diagnostic tool \textcolor{black}{with which to assess} the existence of young and intermediate-age stars \citep{2013ApJ...767...90K,2014ApJ...791..134K, 2016ApJ...820..132K}.
The mid-IR colo\textcolor{black}{u}r of the BCG is 1.197, which is at least \textcolor{black}{one} magnitude bluer than other member galaxies. Thus\textcolor{black}{,} we could conclude that the contribution of the young stellar population within the BCG of J1054 is negligible. 
Considering the nearest neighbo\textcolor{black}{u}r, the BCG of J1054 seems to have undergone dry merging without recent significant star formation. 

\subsection{Member galaxies properties}

\begin{figure*}
\centering
\includegraphics[angle=90,trim={4cm 0 4cm 0},clip,width=2.0\columnwidth]{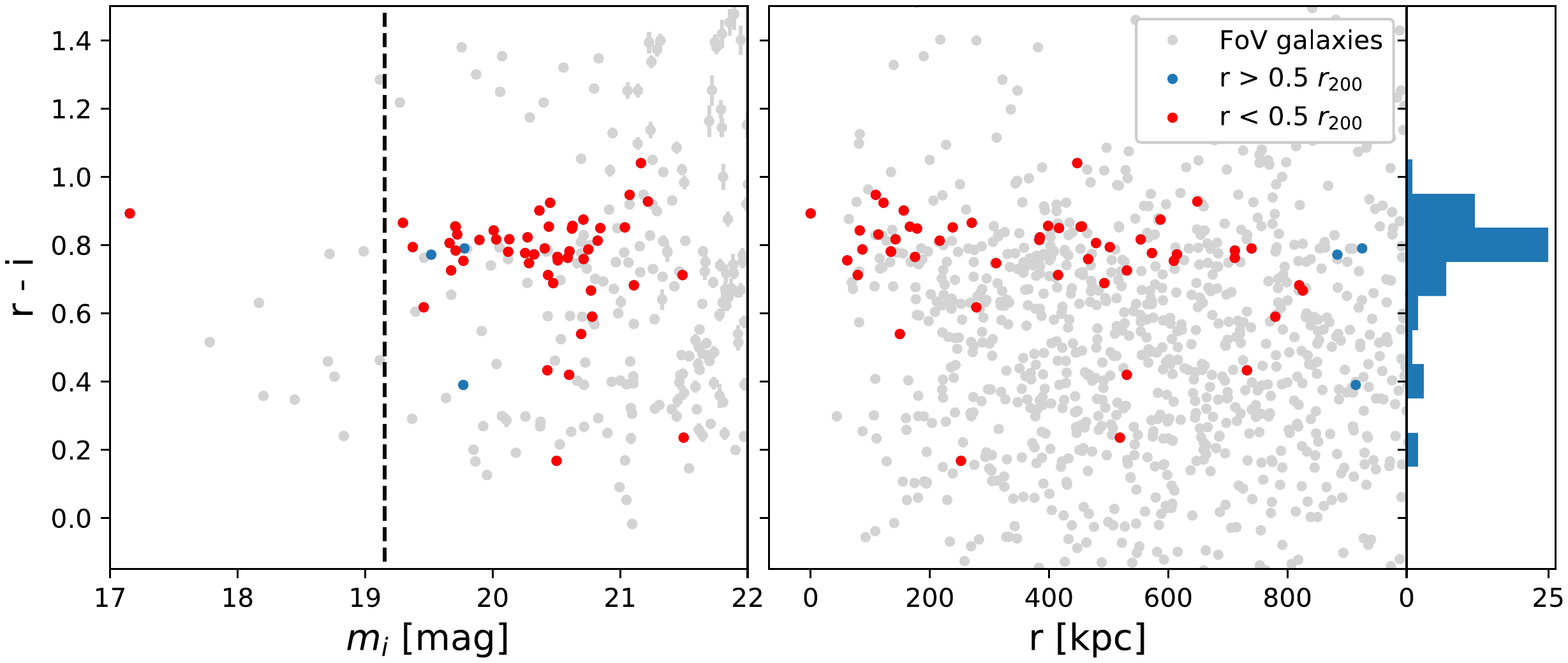}
\caption{(Left) \textit{r}--\textit{i} colour vs. \textcolor{black}{the} \textit{i}--band magnitude for all galaxies (gr\textcolor{black}{e}y dots) within the Gemini field of view in the J1054 field. Among the 73 member galaxies, 53 are detected within the Gemini field of view and \textcolor{black}{are indicated here by the} red and blue dots. Red dots are members within 0.5 $r_{200}$ \textcolor{black}{(}840 kpc\textcolor{black}{)} from the BCG centre, while blue dots are members outside \textcolor{black}{of} 0.5 $r_{200}$. The dashed line is where the  magnitude is $m_{BCG} +2$, showing that J1054 fulfils the criteria to be \textcolor{black}{recognised as} a fossil cluster. The \textit{r}--\textit{i} colour errors are \textcolor{black}{indicated by the grey} errorbar. (Right) \textit{r}--\textit{i} colour distribution for 53 members and field of view galaxies as a function of \textcolor{black}{the} radial distance from the BCG. The histogram on the right side represents the colour distribution of \textcolor{black}{the} 53 member galaxies. The BCG and red sequence galaxies of J1054 have \textit{r}--\textit{i} colour \textcolor{black}{values} of 0.7 $\sim$ 0.9, and \textcolor{black}{the} overall colour \textcolor{black}{values} of \textcolor{black}{the} member galaxies \textcolor{black}{range} widely from 0.17 to 1.04.\label{fig:f10}}
\end{figure*}

The colour and spatial distribution of cluster galaxies could help us to constrain the ICL formation mechanism. We identified 73 cluster member galaxies of J1054 in Section 4.2.1. Among the member galaxies, 53 galaxies are detected within the full field of view of the Gemini images. In Figure \ref{fig:f10}, the colour-magnitude diagram of the member galaxies confirms again that J1054 fulfils the definition of a fossil cluster with a magnitude gap of 2.14 between the BCG and the second brightest galaxy within half \textcolor{black}{of} the projected virial radius ($\sim r_{200}$ = 1.68 Mpc). The actual second brightest member galaxy (magnitude gap from the BCG is 1.56 in \textcolor{black}{the} \texttt{modelMag} from SDSS DR16) is 1271 kpc away from the BCG.

The colour distribution of the member galaxies shows a narrow red sequence \textcolor{black}{with an} \textit{r}--\textit{i} colour of $\sim$ 0.8, and a wide range of  blue galaxies to $\sim$ 0.2 (Figure \ref{fig:f10}). As expected \textcolor{black}{from} a dynamically relaxed cluster, bright red galaxies are dominant in the central region, and faint blue galaxies tend to be located in the outer region of the cluster. 

\begin{figure*}
\centering
\includegraphics[width=2.0\columnwidth]{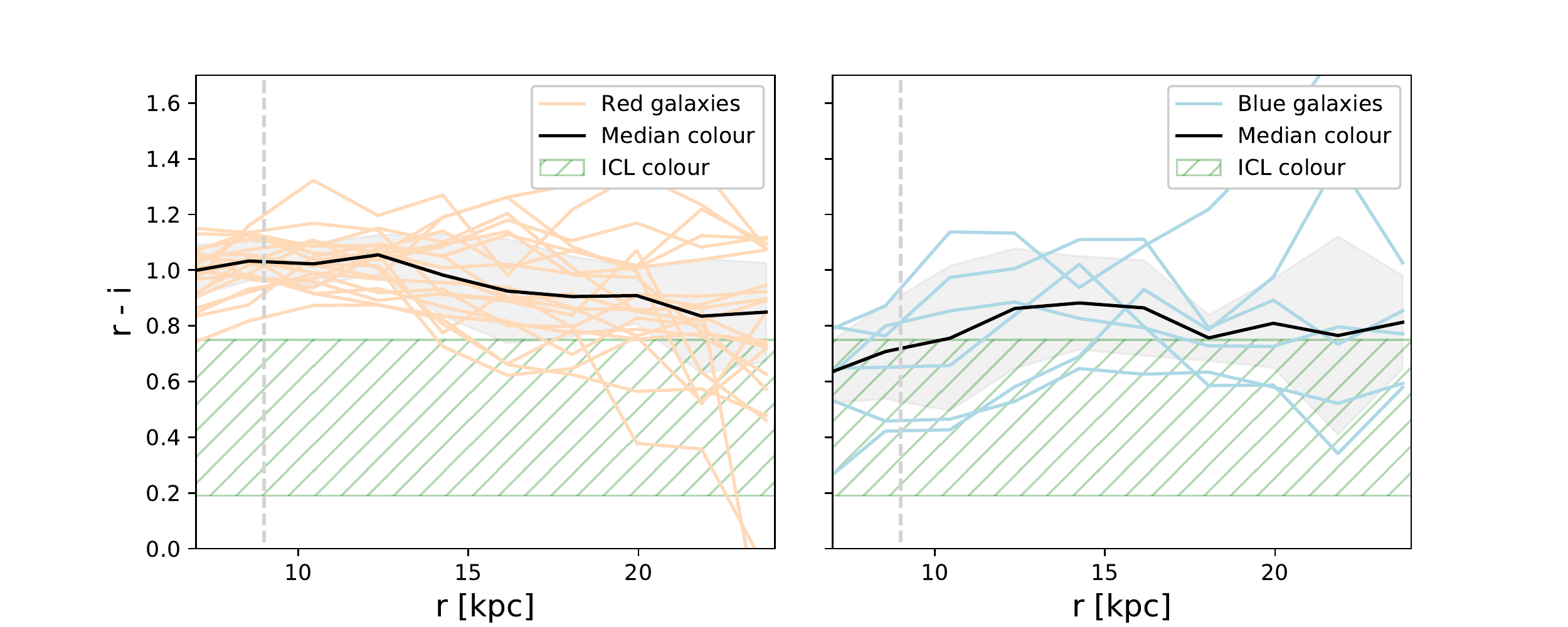}
\caption{The \textit{r}--\textit{i} colour profiles of 21 red sequence galaxies (\textcolor{black}{l}eft) and \textcolor{black}{six} blue galaxies (\textcolor{black}{r}ight) are plotted. The black solid lines represent the median colour profiles. The 1$\sigma$ scatter \textcolor{black}{outcomes} of the profiles are plotted as \textcolor{black}{the} filled gr\textcolor{black}{e}y region in both panel\textcolor{black}{s} (we clipped \textcolor{black}{the} 2$\sigma$ outlier data points due to the sparsity of the sample \textcolor{black}{set}). We consider only \textcolor{black}{points} outside of the radius \textcolor{black}{of approximately} 1.5$\arcsec$ (gr\textcolor{black}{e}y dashed line) to avoid the seeing effect. For comparison, the \textit{r}--\textit{i} colour\textcolor{black}{s} of ICL (\textcolor{black}{0.47 $\pm$ 0.28} in the range of 80 $\sim$ 130 kpc) are indicated (\textcolor{black}{green} hatched region) in both panel\textcolor{black}{s}. \label{fig:f12_2}}
\end{figure*}

We also check\textcolor{black}{ed} the \textit{r}--\textit{i} colour radial profile of 27 member galaxies (except the BCG) in our full\textcolor{black}{-}exposure image area. \textcolor{black}{First, w}e classify 21 red galaxies (\textit{r}--\textit{i} $\geq$ 0.75) and \textcolor{black}{six} blue galaxies (\textit{r}--\textit{i} $<$ 0.75). The magnitude range\textcolor{black}{s} for the red and blue galaxies \textcolor{black}{are} $19.30 < m_i < 21.28$ and $19.47 < m_i < 20.63$, respectively. Figure \ref{fig:f12_2} shows the median colour profile\textcolor{black}{s} of \textcolor{black}{the} red galaxies (left panel) and \textcolor{black}{the} blue galaxies (right panel). The 1 $\sigma$ scatter \textcolor{black}{in} the profiles \textcolor{black}{is} plotted \textcolor{black}{in each case} as \textcolor{black}{the} filled gr\textcolor{black}{e}y region in each panel. In the same manner as the BCG radial profile, we avoid the central region \textcolor{black}{of approximately} 1.5$\arcsec$ ($\sim$ 9 kpc) which \textcolor{black}{is affected by} the seeing effect. Compar\textcolor{black}{ed} to the ICL colour\textcolor{black}{,} which is \textcolor{black}{0.47 $\pm$ 0.28} in the radial range \textcolor{black}{of} 80 $\sim$ 130 kpc, the colour of \textcolor{black}{the} ICL \textcolor{black}{appears} to be related to the outskirt\textcolor{black}{s} (r $>$ 20 kpc) of \textcolor{black}{the} red/blue members and the core of \textcolor{black}{the} blue members.
\section{Discussion} \label{sec:discussion}
Based on the results from the ICL measurements and physical properties of the cluster, we discuss how the ICL is connected to the BCG, member galaxies and the host galaxy cluster.\\

\subsection{What can the colour of ICL tell about the origin of ICL?}
It is well known that the distribution of cluster galaxies in \textcolor{black}{an} optical colour-magnitude diagram is bimodal; quiescent, bright galaxies populate a narrow red sequence and star-forming, faint galaxies form a wide blue cloud \citep[e.g.,][]{2001AJ....122.1861S, 2003ApJ...594..186B, 2004ApJ...600..681B, 2004MNRAS.348.1355B, 2007ApJ...658..884C, 2013ApJ...767...90K}. The ICL colour measurement compared to \textcolor{black}{the} colour distribution of cluster galaxies thus allows \textcolor{black}{us} to constrain the progenitor galaxies of the intracluster stars. 

In the range of 80 $\sim$ 130 kpc, the ICL colour is \textit{r}--\textit{i} = \textcolor{black}{0.47 $\pm$ 0.28}, where the BCG-dominated total light colour is \textit{r}--\textit{i} = \textcolor{black}{0.68 $\pm$ 0.05} (Figure \ref{fig:f6}). \textcolor{black}{Moreover, the colour of the BCG is even redder (\textit{r}--\textit{i} = 0.89 $\pm$ 0.04), which is consistent with the estimated age ($\sim$ 7 $\pm$ 0.5 Gyr) from the BCG spectrum.}
This colo\textcolor{black}{u}r difference suggests that the main component of the ICL differ\textcolor{black}{s} from that of the BCG and bright red galaxies. 
Rather, the colour of the ICL is similar to the outskirts of the red and/or blue member galaxies or the core of the \textcolor{black}{bluest} member (Figure \ref{fig:f12_2}). Furthermore, the colour gradient of the ICL \textcolor{black}{appears to be} negatively steeper than \textcolor{black}{the} total light \textcolor{black}{with the large errors}. This \textcolor{black}{might} indicate that intracluster stars in the central region are older and/or richer \textcolor{black}{in metals} than those in the outer part. This is consistent with previous ICL studies o\textcolor{black}{f} similar redshifts \citep[$z \sim 0.45$;][]{2014A&A...565A.126P, 2015MNRAS.448.1162D, 2018MNRAS.474.3009D}. \cite{2014A&A...565A.126P} reported \textcolor{black}{that} the ICL colour is similar to the envelope of the BCG rather than \textcolor{black}{to the corresponding} centre, and \cite{2018MNRAS.474.3009D} reported \textcolor{black}{that} the ICL colour within 100 kpc is consistent with red sequence galaxies but \textcolor{black}{that it} also has \textcolor{black}{a} negative gradient \textcolor{black}{at} 53 $\sim$ 100 kpc.

These results favour mechanisms \textcolor{black}{according to which} the ICL is not generated during the BCG major merger\textcolor{black}{s} at early times but \textcolor{black}{are} rather gradually accreted by tidal stripping of satellite/infalling galaxies. Similarly, a previous ICL study of the \textcolor{black}{V}irgo cluster \citep{2017ApJ...834...16M} reported the colour of a diffuse plume structure being \textit{B}--\textit{V} = 0.5 $\sim$ 0.6, which was explained as young stars being stripped from the disk of the galaxy or induced recent star formation in the tidal feature.
\textcolor{black}{It is also necessary} to consider the colour evolution of the ICL as well. \cite{2019ApJ...871...24C} predicted \textcolor{black}{that} the ICL colour has a redshift evolution, i.e.\textcolor{black}{,} \textit{B}--\textit{V} = 0.77 at $z=0$ and 0.63 at $z=1$. \textcolor{black}{Moreover}, several observational studies have shown that the ICL \textcolor{black}{during} intermediate redshifts is younger than \textcolor{black}{those} nearby \citep{2011MNRAS.414..602T, 2014ApJ...794..137M, 2018MNRAS.474..917M, 2016A&A...592A...7A, 2017ApJ...846..139M, 2018ApJ...857...79J}.
Therefore, the overall bluer \textit{r}--\textit{i} (\textit{B}--\textit{V} in the rest-frame) colour of the ICL at $z=0.47$, compared to the Virgo cluster \citep[\textit{B}--\textit{V} = 0.7 $\sim$ 1.0;][]{2010ApJ...720..569R, 2017ApJ...834...16M}, is comprehensible.

\subsection{Does the spatial distribution of ICL follow that of the cluster galaxies?}

\begin{figure*}
\centering
\includegraphics[height=2.75in,trim={0cm 6cm 0cm 6cm},clip]{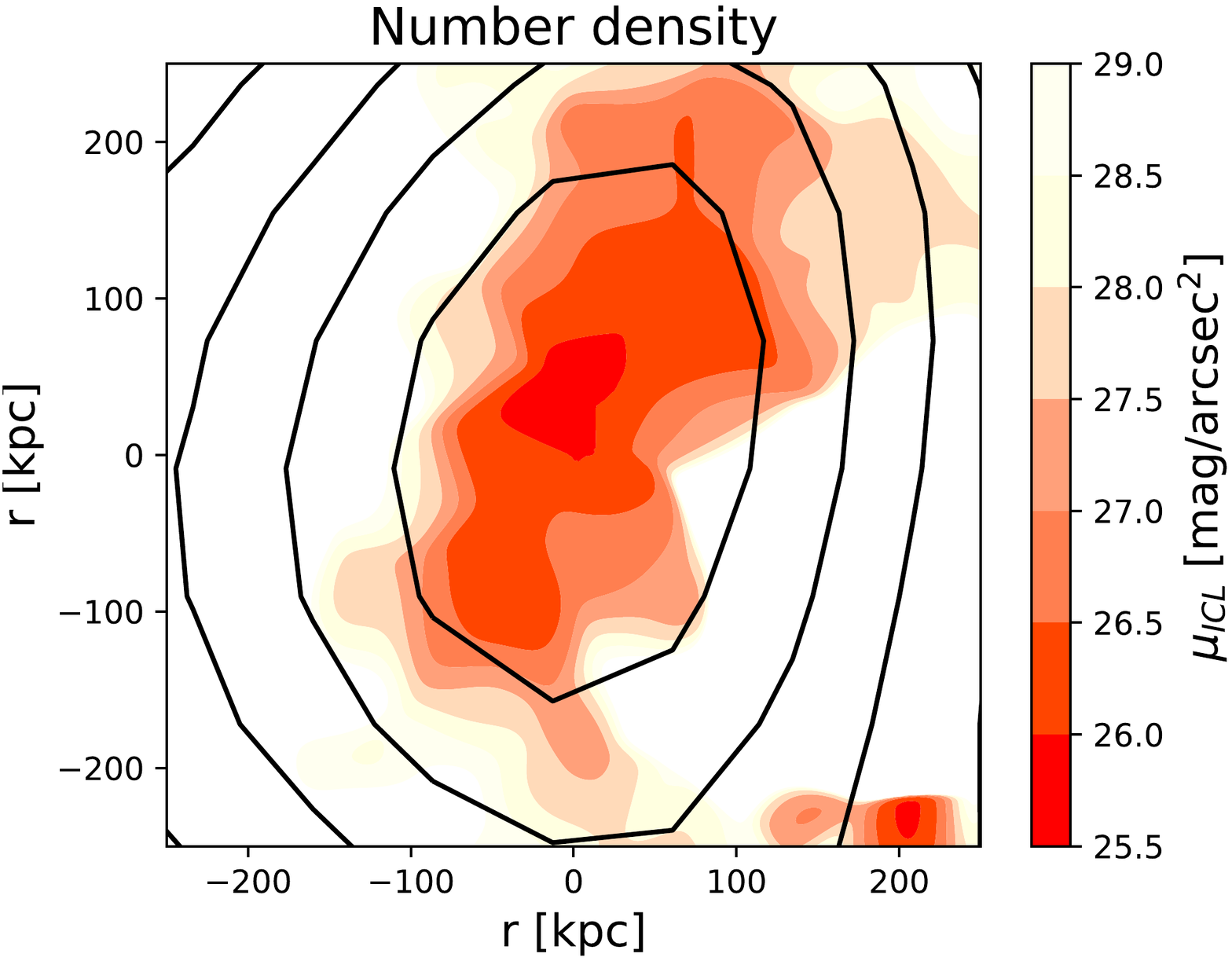}
\includegraphics[height=2.75in,trim={0cm 6cm 0cm 6cm},clip]{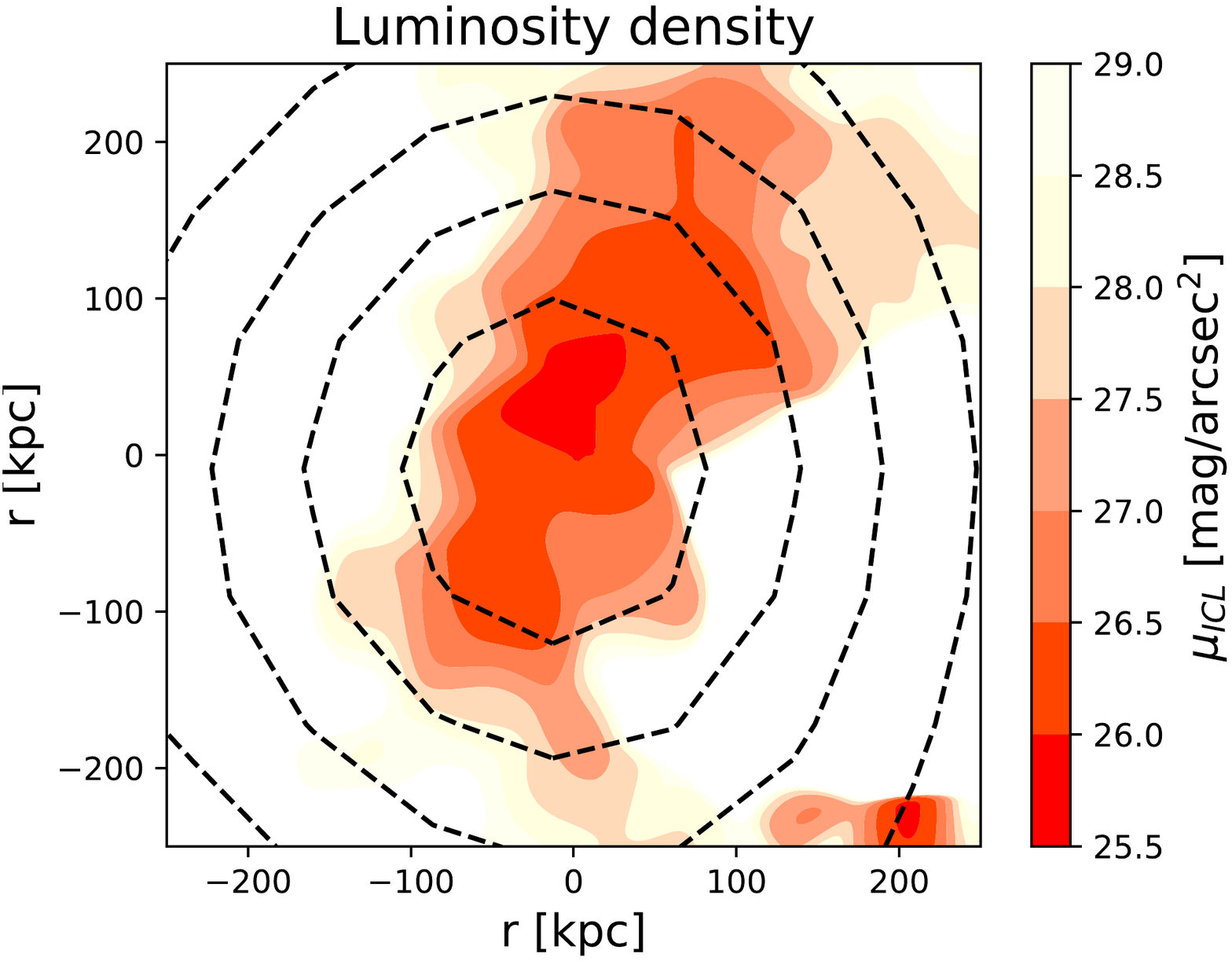}
\caption{Contour of the galaxy number density map (\textcolor{black}{l}eft, black solid line) and contour of the luminosity\textcolor{black}{-}weighted number density map (\textcolor{black}{r}ight, black dashed line), over\textcolor{black}{-}plotted on the \textit{i}--band diffuse light surface brightness 2D map. The 73 member galaxies are selected using \textcolor{black}{the} Caustic method (see Section 4.2.1). Their spatial distribution results \textcolor{black}{in} the number density map, and \textcolor{black}{assign}ing weight \textcolor{black}{to} it using their \textit{i}--band luminosity results \textcolor{black}{in} the luminosity density map. \textcolor{black}{Because} J1054 is a BCG\textcolor{black}{-}dominated fossil cluster, we associate the number density map with the member\textcolor{black}{-}galaxy\textcolor{black}{-}oriented ICL formation scenario, and the luminosity density map with the BCG\textcolor{black}{-}oriented scenario. \textcolor{black}{Despite the fact that} the number density map is generated using the \textcolor{black}{location data of} member galaxies in \textcolor{black}{the} SDSS catalog\textcolor{black}{ue}, which is independent \textcolor{black}{of} our observational image data, the shape coincides with the diffuse light distribution. The measured ellipticities of the \textcolor{black}{three} innermost contours are 0.71, 0.66, \textcolor{black}{and} 0.64 for the number density, and 0.0, 0.49, \textcolor{black}{and} 0.49 for the luminosity density. The position angle is measured in degrees counter-clockwise from the vertical axis of the image (see Section 5.2 for detail\textcolor{black}{s}). \label{fig:f13}}
\end{figure*}

The spatial distribution of cluster member galaxies can be visualised as a galaxy number density map. Using the additional luminosity information (\textit{i}--band magnitudes from SDSS) of each galaxy gives us a galaxy luminosity density map (Figure \ref{fig:f13}). We generate the 2D projected density maps on a 30 $\times$ 30 pixels binned grid over the spatial range where the 73 member galaxies reside and \textcolor{black}{conduct} Gaussian smooth\textcolor{black}{ing of} each field with a factor of 2.2.

In Figure \ref{fig:f13}, the contours for the number density (derived from \textcolor{black}{the} positions of \textcolor{black}{the} galaxies) are elongated, whereas those for the luminosity\textcolor{black}{-}weighted density are rather more circular. The measured ellipticities of the \textcolor{black}{three} innermost contours are 0.71 \textcolor{black}{$\pm$0.09}, 0.66 \textcolor{black}{$\pm$0.05}, \textcolor{black}{and} 0.64 \textcolor{black}{$\pm$0.04} for the number density and 0.0 \textcolor{black}{$\pm$0.08}, 0.49 \textcolor{black}{$\pm$0.18}, \textcolor{black}{and} 0.49 \textcolor{black}{$\pm$0.10} for the luminosity density. \textcolor{black}{Regarding} the orientation, the measured position angles (in degrees counter-clockwise from the vertical axis of the image) of the \textcolor{black}{three} innermost contours are -75.10 \textcolor{black}{$\pm$ 9.74}\degr, -70.51 \textcolor{black}{$\pm$ 7.45}\degr, \textcolor{black}{and} -81.40 \textcolor{black}{$\pm$ 38.96}\degr\ for the number density, and 44.71 \textcolor{black}{$\pm$ 10.31}\degr, 63.06 \textcolor{black}{$\pm$ 42.40}\degr, \textcolor{black}{and} -82.55 \textcolor{black}{$\pm$ 40.11}\degr\ for the luminosity density. \textcolor{black}{Because} the luminosity density map is derived from \textcolor{black}{the} positions and luminosities of the cluster member galaxies, it does not take into account the shape of the BCG itself. The ellipticity of the BCG is 0.26\textcolor{black}{4} \textcolor{black}{$\pm$ 0.001} and the \textcolor{black}{corresponding} position angle is -18.33 \textcolor{black}{$\pm$ 0.09}\degr, \textcolor{black}{providing} better \textcolor{black}{agreement} with the luminosity density map than with the number density map.

In the luminosity density contours, which draw concentric circles (right panel of Figure \ref{fig:f13})\textcolor{black}{,} we can clearly see J1054's character\textcolor{black}{istic} as a fossil cluster with dominating BCG. The number density map is closely linked to \textcolor{black}{the} cluster member galaxies, whereas the luminosity density map is more related to the luminosity\textcolor{black}{-}dominating BCG. Comparing the number and luminosity density maps with the ICL map \textcolor{black}{may provide} a hint to the origin of ICL. Regarding the elongated \textcolor{black}{(ellipticity of 0.95 $\pm$ 0.02)} and tilted \textcolor{black}{(position angle of -74.20 $\pm$ 1.08\degr)} shape, the ICL map is \textcolor{black}{better} matched \textcolor{black}{(overlap at the 2 $\sigma$ level) to} the galaxy number density map rather than \textcolor{black}{to} the luminosity\textcolor{black}{-}weighted density map, which implies that the ICL distribution is related \textcolor{black}{more to} cluster galaxies than solely \textcolor{black}{to} the BCG.
We note that the outer region of cluster galaxies, if not adequately masked, may artificially induce a spatial correlation between the member galaxies and the ICL. In order to mitigate this risk\textcolor{black}{,} we masked luminous objects \textcolor{black}{to} measur\textcolor{black}{e} the ICL in a conservative manner. 

\subsection{How abundant is the ICL of J1054 compared to \textcolor{black}{that of} other galaxy clusters?}

\begin{figure*}
\centering
\includegraphics[angle=90,trim={5cm 0cm 5cm 0},clip,width=2.0\columnwidth]{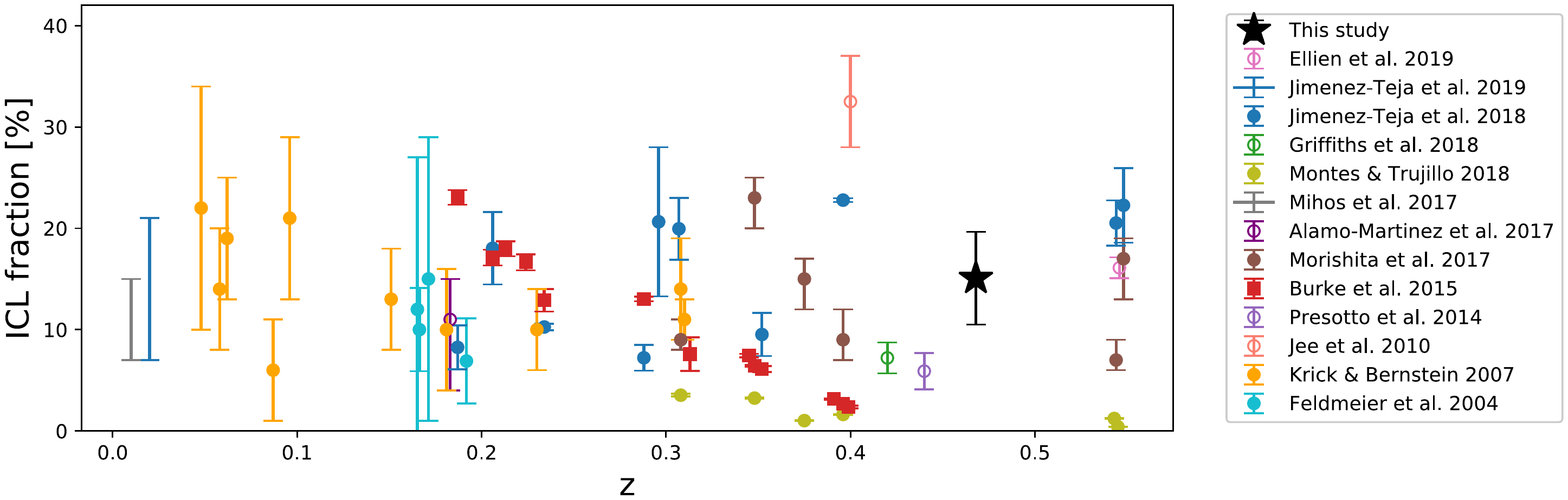}
\caption{Reference ICL fraction\textcolor{black}{s} of galaxy clusters in various redshifts using various ICL detection technique\textcolor{black}{s}. The ICL fraction of J1054 ($15.07 \pm 4.57 \%$) is \textcolor{black}{indicated by the} black star. \textcolor{black}{Because these} results are derived through different ICL defining method\textcolor{black}{s}, \textcolor{black}{a} comparison between ICL fractions should be \textcolor{black}{done} with caution. The results from Burke et al. (marked as red squares) are derived from surface brightness cut\textcolor{black}{s} as conducted in this study (see detail\textcolor{black}{s} in Section 5.3)\label{fig:f14}.}
\end{figure*}

Investigating the ICL fraction allows us a further important constraint on the ICL formation and evolution \textcolor{black}{processes}. If intracluster stars in nearby galaxy clusters have been stripped from the cluster galaxies through numerous galaxy interactions, we \textcolor{black}{can} predict that the ICL fraction over a sample of galaxy clusters is \textcolor{black}{close}ly related \textcolor{black}{to} redshift and dynamic state. Recent simulations show a strong evolution of the ICL fraction with \textcolor{black}{the} redshift \citep{2011ApJ...732...48R, 2014MNRAS.437.3787C, 2015MNRAS.451.2703C}. For example, \cite{2011ApJ...732...48R} predicted \textcolor{black}{via a} simulation that the ICL fraction increases from $6.0 \sim 8.5\%$ at $z=0.5$ to $9.5 \sim 13.2\%$ at $z=0$, using a surface brightness threshold of $\mu_V$ = 26.5 mag/arcsec$^2$. 

However, observational studies \textcolor{black}{overall} do not \textcolor{black}{yet appear} to support the\textcolor{black}{se} simulation results. Figure \ref{fig:f14} shows our ICL fraction measured in J1054 along with those in other clusters with varying redshift\textcolor{black}{s}. 
There \textcolor{black}{does not appear to be a} clear trend between the ICL fraction and \textcolor{black}{the} redshift. However, there are several ambiguities present. First, a common ICL detection method \textcolor{black}{is lacking}. The methods known \textcolor{black}{thus} far are as follows. \cite{2014A&A...565A.126P}, \cite{2017ApJ...849....6A}, \cite{2017ApJ...846..139M} and \cite{2018MNRAS.475.2853G} modelled galaxies using \texttt{GALFIT} and used this model to subtract the BGC and galaxy light contributions. They defined \textcolor{black}{the} ICL as the residual light after the removal of all light contribution\textcolor{black}{s} of galaxies. \cite{2016A&A...592A...7A} and \cite{2019A&A...628A..34E} used a wavelet-based method to model the surface brightness distribution of all galaxies\textcolor{black}{,} subtract\textcolor{black}{ing} them to obtain the residual ICL map. \cite{2018ApJ...857...79J, 2019A&A...622A.183J} used the CICLE method which is similar to the wavelet method. \cite{2004ApJ...609..617F}, \cite{2007AJ....134..466K}, \cite{2010ApJ...717..420J}, \textcolor{black}{\cite{2012MNRAS.425.2058B, 2015MNRAS.449.2353B}} \cite{2017ApJ...834...16M}, \textcolor{black}{and} \cite{2018MNRAS.474..917M} and this study use \textcolor{black}{the} surface brightness threshold. For example, \cite{2015MNRAS.449.2353B} (as \textcolor{black}{indicated by the} red squares in Figure \ref{fig:f14}) show a clear tendency \textcolor{black}{of} the ICL fraction increas\textcolor{black}{ing} with \textcolor{black}{a} decreas\textcolor{black}{e in the} redshift, whereas the data of Jimenez-Teja et al. and Morishita et al. (\textcolor{black}{indicated by the} filled circles) do not show any relation between the ICL fraction and \textcolor{black}{the} redshift. 
Second, the measurement technique for the ICL fraction (i.e. how to quantify the total cluster and ICL light) is different in each study. \cite{2018MNRAS.474..917M} \textcolor{black}{demonstrated} how the different definitions of \textcolor{black}{the} ICL affect the ICL fraction. Thus, it is difficult directly \textcolor{black}{to} compare multiple observational studies. To address this issue, we \textcolor{black}{attempt} to compare the result\textcolor{black}{s from the} method adopted in this study.
Compar\textcolor{black}{ed to} Burke et al. (the red square symbols), wh\textcolor{black}{ere a} surface brightness threshold technique \textcolor{black}{was used} for the ICL measurement, J1054 \textcolor{black}{appears} to have a higher ICL fraction ($15.07 \pm 4.57 \%$ in the \textit{i}--band which corresponds to \textcolor{black}{a} \textit{V}--band in the rest-frame, black star symbol in Figure \ref{fig:f14}). Even if we take the lower limit ($14.50 \pm 4.40 \%$ in the \textit{i}--band, calculated in Section 4.1), the ICL fraction of J1054 is still higher than that of \textcolor{black}{the} clusters in Burke et al.

A direct comparison of the ICL fraction measurement method of Burke et al. with our data is not possible, \textcolor{black}{as} their lower limit of \textcolor{black}{the} surface brightness is 26.4  mag/arcsec$^2$ (in \textcolor{black}{the} F625W filter), which is brighter than the ICL surface brightness threshold \textcolor{black}{of} 26.67  mag/arcsec$^2$ in the \textit{r}--band (converted from 25  mag/arcsec$^2$ \textcolor{black}{in the} rest-frame \textit{B}--band). If we take our \textit{r}--band detection limit \textcolor{black}{of} 29.1  mag/arcsec$^2$ as the lower limit, adopting their radius limit ($R_{500}$) and their point source/non-cluster galax\textcolor{black}{y} masking, \textcolor{black}{which was enlarged by 1.5 times,} the calculated ICL fraction (ICL over \textcolor{black}{the} total cluster light of BCG + ICL + satellite galaxies) of J1054 becomes even higher, \textcolor{black}{at} 19.77\%,
due to our deeper detection limit. Note that, as Burke et al. has already pointed out, their ICL fraction var\textcolor{black}{ies} significantly with the lower limit of the surface brightness. \cite{2011ApJ...732...48R} simulated the redshift evolution of the ICL fraction for various surface brightness thresholds (see their Figure 4) and predicted that the contribution of ICL dimmer than $\mu_{V} \sim$ 26.5 mag/arcsec$^2$ at $z\sim 0.47$ is about $\sim$ 8.5\%. \textcolor{black}{Because the lower limit of} their simulated surface brightness is much deeper than current observations allow (reaching \textcolor{black}{as low as} $\mu_{V} =$ 35 mag/arcsec$^2$), we may regard the ICL fraction \textcolor{black}{of} $\sim$ 8.5\% as the amount of missing ICL due to the shallower detection limit \textcolor{black}{relative to} that surface brightness level. Thus, if our detection limit were as shallow as that of Burke et al., we would have missed that corresponding amount of ICL as well. If we accept this assumption, the estimated ICL fraction of J1054 according to the method \textcolor{black}{by} Burke et al. decreases to $\sim$ 11.27\%, which still \textcolor{black}{appears to be} high for their ICL fraction-redshift relation.

Abell 2261 is one of the sample clusters of Burke et al., which is also a massive fossil cluster with $M_{vir} \sim 1.76 \times 10^{15} M_{\odot}$\citep{2015ApJ...806....4M}. Considering its ICL fraction of $16.64 \pm 0.78\%$ \citep{2015MNRAS.449.2353B} at $z=0.22$ and the converted ICL fraction of J1054, \textcolor{black}{of} $\sim$ 11.27\% at $z=0.47$, the growth factor over the corresponding redshift range is $\sim$ 1.48, which is likely to be consistent with simulation studies \citep{2011ApJ...732...48R, 2014MNRAS.437.3787C}. Rudick et al. predicted a factor of $\sim$ 1.35 and Contini et al. predicted factors of $\sim$ 1.47 and $\sim$ 1.56 for a tidal model with and without a merger channel, respectively. Among the various ICL growth mechanism\textcolor{black}{s} simulated by Contini et al. (see their Figure 6), the model with constant stripping and \textcolor{black}{a} merger channel (i.e., both the main mechanisms at play) matches our estimation of ICL growth \textcolor{black}{best}.

Assuming that fossil clusters at $0.4 < z < 0.5$ have BCGs whose formation\textcolor{black}{s} ha\textcolor{black}{ve} already finished, and further positing that \textcolor{black}{a} BCG major merger is the main ICL production mechanism, this would imply that most of the ICL would have been generated up until that epoch.
From this point of view, our high ICL fraction could support the BCG origin scenario, \textcolor{black}{whereas} it is also an expected result from a fossil cluster, which is a dynamically mature system where galaxy-galaxy interactions such as tidal stripping have already taken place. 
Moreover, the bluer ICL colour compared to the BCG and the spatial distribution of the ICL aligned with member galaxies strengthen the possibility that J1054 has a fraction of \textcolor{black}{its} ICL component generated by stripping processes. 

\section{Conclusions} \label{sec:conclusion}
We have presented \textcolor{black}{an} ICL study of J1054, a massive fossil cluster at $z=0.47$, to improve constraints on the ICL formation mechanism. Gemini deep imaging and MOS observations were conducted to detect the diffuse ICL feature\textcolor{black}{s} of J1054 and investigate its connection with the BCG and cluster galaxies. Our imaging data reach to a very low surface brightness (29.1 and 28.7 mag/arcsec$^2$ for \textcolor{black}{the} \textit{r} and \textit{i}--band, respectively)\textcolor{black}{, sufficient for} explor\textcolor{black}{ations of} the ICL feature out to 155 kpc.

The ICL colour is bluer than the BCG and bright, red cluster galaxies beyond 70 kpc from the location of the BCG\textcolor{black}{,} and its colour gradient \textcolor{black}{appears to be} steeper than the total light \textcolor{black}{with the large errors} (Figure \ref{fig:f6}). The blue colour (\textit{r}--\textit{i} = \textcolor{black}{0.47 $\pm$ 0.28}) of the ICL in the outer region (80 $\sim$ 130 kpc) disfavour\textcolor{black}{s} the ICL formation mechanism related to major mergers\textcolor{black}{,} including \textcolor{black}{those associated with} the BCG. Instead, our findings \textcolor{black}{may} support mechanisms produced by \textcolor{black}{the} recent stripping of the outskirts of infalling/satellite galaxies.

When comparing the spatial distribution of the ICL with cluster galaxies, we find that it is more likely to match the number density map rather than \textcolor{black}{the} luminosity\textcolor{black}{-}weighted density map (Figure \ref{fig:f13}), which implies that \textcolor{black}{the} ICL distribution is related \textcolor{black}{more to} member galaxies than solely \textcolor{black}{to} the BCG.


We estimate an ICL fraction of $15.07 \pm 4.57 \%$ in the range of $60 \sim 155$ kpc, which \textcolor{black}{appears} higher than the ICL fraction-redshift trend reported by Burke et al. (Figure \ref{fig:f14}). Converting the ICL fraction of J1054 to match the detection limit \textcolor{black}{in} Burke et al. and comparing it with the ICL fraction of \textcolor{black}{a} massive fossil cluster sample, Abell 2261\textcolor{black}{,} at $z=0.22$, the estimated growth factor for this redshift interval \textcolor{black}{appears} to be consistent with the tidal stripping $+$ merger model of \cite{2014MNRAS.437.3787C}.

\textcolor{black}{Describing t}he ICL formation process of J1054 through a single mechanism \textcolor{black}{may} not be sufficient. Ideally\textcolor{black}{,} we would require a sample of massive fossil clusters with varying redshift\textcolor{black}{s} to draw a firm conclusion \textcolor{black}{about} the dominant ICL formation mechanism during the hierarchical growth of the cluster. Obtaining high-z ($z \geq 1$) fossil cluster samples in future surveys\textcolor{black}{, such as the Vera C. Rubin Observatory's Legacy Survey of Space and Time (LSST)} would be particularly helpful regarding the early BCG build-up and late ($z < 1$) tidal stripping ICL production scenario. Furthermore, \textcolor{black}{a} detailed analysis on the ICL colour, spatial distribution and fraction in the sample \textcolor{black}{will} help to constrain the model \textcolor{black}{to} the origin of the ICL. 

\section*{Acknowledgements}
We thank the anonymous referee for useful comments that have greatly improved this paper.
We thank Cristiano Sabiu and Kwangil Seon for helpful discussions. We thank Ana Laura Serra for allowing us to use the Caustic App and we thank Hoseong Hwang and Hyunmi Song for help using the Caustic method. We thank Woowon Byun for help estimating the BCG age. We thank Hogyu Lee and Sungchul Yang for conducting the Gemini (GMOS-N) observation. Based on observations obtained at the Gemini Observatory, acquired through the Gemini Observatory Archive and processed using the Gemini IRAF package, which is operated by the Association of Universities for Research in Astronomy, Inc., under a cooperative agreement with the NSF on be of the Gemini partnership: the National Science Foundation (United States), National Research Council (Canada), CONICYT (Chile), Ministerio de Ciencia, Tecnolog\'{i}a e Innovaci\'{o}n Productiva (Argentina), Minist\'{e}rio da Ci\^{e}ncia, Tecnologia e Inova\c{c}\~{a}o (Brazil), and Korea Astronomy and Space Science Institute (Republic of Korea). JWK acknowledges support from the National Research Foundation of Korea (NRF), grant No. NRF-2019R1C1C1002796, funded by the Korean government (MSIT). 

\section*{Data Availability}
The data underlying this article are available in the Gemini Observatory Archive at https://archive.gemini.edu/searchform, and can be accessed with Program ID:GN-2018A-Q-201/PI:Jaewon Yoo.
 



\bibliographystyle{mnras}
\bibliography{main} 








\bsp	
\label{lastpage}
\end{document}